\documentclass[aps,pra,twocolumn,floatfix,10pt,showpacs]{revtex4-1}
\usepackage{graphicx}
\usepackage{amssymb,amsmath}
\usepackage{color}
\usepackage{bbold}
\usepackage{tabularx}

\newcommand{\ket}[1]{|#1\rangle}				
\newcommand{\bra}[1]{\langle #1|}				
\newcommand{\ew}[1]{\langle #1 \rangle}				
\newcommand{\ketbra}[1]{| #1 \rangle \langle #1 |}		
\def\cdott{\!\cdot\!}
\newcommand{\scpr}[2]{\left( {#1} \cdott {#2} \right) }
\newcommand{\scprt}[2]{\left( {#1} \cdott \mathbb{1}_\perp \cdott {#2} \right) }
\renewcommand{\vec}[1]{\boldsymbol{#1}}
\def\aout{a_{\textrm{out}}}
\def\ain{a_{\textrm{in}}}
\def\kR{\kappa_{R}}
\def\dc{\Delta_{C}}

\allowdisplaybreaks[2]

\begin{document}
\title{Collective effects between multiple nuclear ensembles in an x-ray cavity-QED setup}

\author{Kilian P. Heeg}
\author{J\"org Evers}
\affiliation{Max-Planck-Institut f\"ur Kernphysik, Saupfercheckweg 1, 69117 Heidelberg, Germany}

\date{\today}
\begin{abstract}
{The setting of M\"ossbauer nuclei embedded in thin-film cavities has facilitated an aspiring platform for x-ray quantum optics as shown in several recent experiments. Here, we generalize the theoretical model of this platform that we developed earlier~[Phys.~Rev.~A~{\bf 88}, 043828 (2013)]. The theory description is extended to cover multiple nuclear ensembles and multiple modes in the cavity. While the extensions separately do not lead to qualitatively new features, their combination gives rise to cooperative effects between the different nuclear ensembles and distinct spectral signatures in the observables. A related experiment by R\"ohlsberger \emph{et al.}~[Nature~{\bf 482}, 199 (2012)] is successfully modeled, the scalings derived with semiclassical methods are reproduced, and a microscopic understanding of the setting is obtained with our quantum mechanical description.}
\end{abstract}


\maketitle

\section{Introduction}
With the advent of novel light sources, the emerging field of x-ray quantum optics has gained considerable momentum, both on the experimental and theoretical side~\cite{Adams2013}. While ideas based on quantum coherence and interference could in principle be realized with inner-shell electrons, solid state targets obeying the M\"ossbauer effect~\cite{Moessbauer1958} have sparked interest in many recent works~\cite{Shvydko1996,Palffy2009,Shakhmuratov2009,Shakhmuratov2011,Liao2012,Liao2012b,Vagizov2014}.

A particularly interesting setting in which quantum optics with M\"ossbauer nuclei can be realized is specifically engineered planar x-ray cavities. Embedding a thin layer of resonant nuclei in such cavities has facilitated the observation of a number of phenomena, such as the cooperative Lamb shift and single-photon superradiance~\cite{Roehlsberger2010}, Fano line shape control and interferometric phase measurements~\cite{Heeg2014b}, magnetically controlled reflection spectra modified by spontaneously generated coherences~\cite{Heeg2013} and group velocity control of x-ray pulses~\cite{Heeg2014}.

However, there is an additional cavity configuration which has sparked interest recently. In a setting with two particularly placed ensembles of M\"ossbauer nuclei in the cavity, the iron isotope $^{57}$Fe with its transition at $14.4$~keV, it was possible to observe a reflection spectrum with a deep interference minimum in the center due to the phenomenon of electromagnetically induced transparency (EIT)~\cite{Harris1997,Fleischhauer2005,Roehlsberger2012}. This is a remarkable result, since typically two coherent driving fields are required for this effect to emerge. In Ref.~\cite{Roehlsberger2012}, however, the EIT experiment was established in a thin-film cavity with only a single excitation from a synchrotron beam, whereas the second field was intrinsically provided by intracavity couplings between the two $^{57}$Fe layers. For the modeling of the experiment, different semiclassical approaches can be employed~\cite{Parratt1954,Roehlsberger2005}, while a consistent description based on quantum optics as usually desired for EIT is still lacking.

A first quantum optical model for the light-matter interaction in the cavity was developed in Ref.~\cite{Heeg2013b}. However, it does not yet cover cavity settings with multiple resonant layers, and hence it is also not yet capable of describing the EIT experiment. But motivated by the expected significance of multilayer configurations, it would be highly desirable to also have a microscopic theory at hand, which allows for a deeper understanding. Topics of interest include the nature of the above-mentioned intrinsic cavity-mediated coupling between the different resonant layers, and perspectives on how multilayer cavities can be specifically engineered. We note that other approaches based on scattering theory were employed to model the two-layer layout~\cite{Xu2013}; however to our knowledge they remained unsuccessful in providing a quantitative description of the EIT experiment.

In this work, we generalize the quantum optical theory from Ref.~\cite{Heeg2013b} with the aim to describe the single-photon EIT experiment~\cite{Roehlsberger2012} and related settings. To this end, we extend the description to include multiple cavity modes as well as multiple layers. The extension to multiple modes allows us to accurately describe the cavity reflection in the absence of resonant nuclei. However, the general shape of the nuclear contribution to the measured signal turns out to be unchanged when considering the two extensions separately. In this case, the line shape predicted in the absence of a magnetic hyperfine splitting is a Lorentz profile, in which only the coefficients are modified due to the additional elements in the theory. However, if multiple cavity modes and multiple layers are considered simultaneously, a new class of nuclear reflection spectra is obtained. In particular, restricting the analysis to two resonant layers, the EIT-like spectrum observed in Ref.~\cite{Roehlsberger2012} is reobtained and the predicted scalings are in accordance with previous semiclassical calculations. Due to the microscopic ansatz of our theoretical model, we can provide a full quantum interpretation of the system. A good agreement to numerical results obtained from semiclassical descriptions is found over a broad parameter range. Furthermore, our extended model opens up avenues to engineer a broader set of effective level schemes at x-ray energies by combining the advanced possibilities of multiple modes and layers together with magnetic hyperfine splitting.

This paper is structured as follows: In Sec.~\ref{sec:basic} we recapitulate the cavity system and the basic model which was developed for its description in Ref.~\cite{Heeg2013b}. After this, we generalize the model to include multiple layers and multiple modes. In Secs.~\ref{sec:effect_modes} and~\ref{sec:effect_layers}, we analyze the effects of the two extensions separately and in Sec.~\ref{sec:effect_both} the consequence of both extensions applied simultaneously are discussed. Finally, the general model is applied to describe the EIT setting from Ref.~\cite{Roehlsberger2012}.

\section{Recap of the basic model\label{sec:basic}}
\begin{figure}[t]
 \centering
 \includegraphics{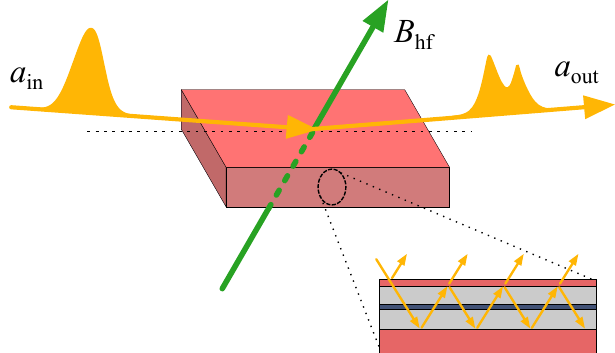}
 \caption{\label{fig:setup}(Color online) Schematic of the considered setup. A thin-film cavity is probed by hard x rays in grazing incidence. The light-matter interaction with the resonant nuclei in the center of the cavity, possibly under the influence of a magnetic field, modifies the observed reflected signal.}
\end{figure}

The setup which is considered in this work is a thin-film cavity with embedded M\"ossbauer nuclei, probed by hard x rays in grazing incidence as visualized in Fig.~\ref{fig:setup}. Such a cavity is formed by a stack of different materials. At its boundaries materials with a high electron density, such as platinum or palladium, act as mirrors, while the material in the center, e.g.~carbon, has a low electron density and provides a guiding layer for the x rays. For certain incident angles in the mrad range, the x rays can resonantly excite a guided cavity mode and propagate inside the cavity, rendering the structure a waveguide-like system. By embedding M\"ossbauer nuclei in the center of the waveguide, a near-resonant interaction of the x-ray light with the transitions of the resonant nuclei is achieved. The reflected signal forms the main observable, and its spectral shape is crucially influenced by the light-matter interaction in the cavity. As proven in a number of recent experiments~\cite{Roehlsberger2010,Roehlsberger2012,Heeg2013,Heeg2014,Heeg2014b}, this setup constitutes an auspicious platform for the exploration of quantum optical phenomena in the x-ray regime.

A quantum optical model for the description of the x-ray light-matter interaction in these thin-film cavities was introduced in Ref.~\cite{Heeg2013b}, however, it is limited to a small subset of possible cavity layouts. In the following, we will briefly present the existing theoretical approach, before we then continue to generalize the model in order to cover more elaborate scenarios.

\subsection{Cavity\label{sec:basic_cavity}}
As already mentioned above, guided cavity modes can be excited for certain resonance angles $\theta_0$. These angles depend on the cavity layout, such as the materials and the layer thicknesses, and can be determined by computing the angular-dependent reflectance in the absence of resonant nuclei. At certain positions of this curve, the reflection is strongly suppressed, indicating the presence of a guided cavity mode. The reason for this suppression is the destructive interference between the reflection directly at the cavity surface and the reflection of the light which entered the cavity mode.

\begin{figure}[t]
 \centering
 \includegraphics{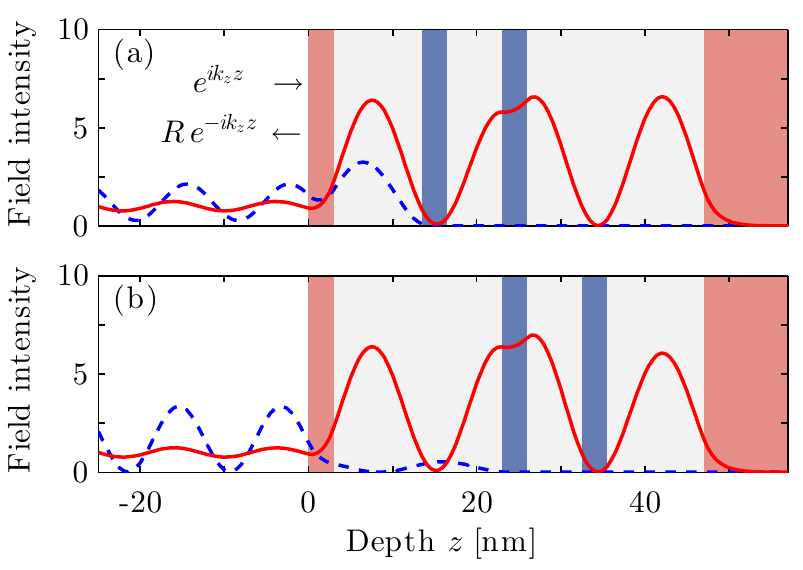}
 \caption{\label{fig:field}(Color online) Field distribution in the two cavities defined in Tab.~\ref{tab:layers_eit} is shown. The incidence angle ($\theta \approx 3.5$~mrad) is chosen such that the third guided mode is driven. The cavity field is normalized to an input field with unity intensity. Solid red lines take into account only electronic scattering in the cavity, which is realized at x-ray frequencies off-resonant to the nuclear transition in $^{57}$Fe, dashed blue curves assume resonant driving of the transition. The shaded areas (red, gray, blue) indicate the positions of the different cavity materials (Pt, C, Fe).}
\end{figure}

Exemplary field distributions for two particular cavities, which will be analyzed later, are shown in Fig.~\ref{fig:field}. The incidence angle is chosen such that the third guided mode is driven, which is reflected in the three antinodes of the field intensity inside the cavity. The external cavity field stems from the interference of the incident (``$\exp{(i k_z z)}$'') and reflected beam (``$R \exp{(-i k_z z)}$''). Hence, the suppression in the reflectance $R \ll 1$, which is characteristic for the guided modes, is visible as a small modulation of the external field. In contrast, if the x-ray frequency matches the transition of the nuclei in the cavity (dashed lines in Fig.~\ref{fig:field}), the relative strength of the reflected light becomes larger as it now consists not only of electronic, but also nuclear scattering contributions which adds up to the observed signal. Note that the external field in Fig.~\ref{fig:field} is shown close to the cavity surface and therefore it an interference pattern of the incident and the reflected light appears. In the far field, however, these two contributions are easily distinguished due to the different propagation direction, as visualized in the schematic in Fig.~\ref{fig:setup}.

Since the different cavity modes are well separated in their resonant incident angles by several mrad, only at most one mode is usually driven near its resonance. Nevertheless, to take into account the two polarization states of the x rays perpendicular to the propagation direction, two cavity modes $a_1$ and $a_2$ are included in the theoretical description. In the rotating frame of the external driving field with frequency $\omega$, the Hamiltonian characterizing the free evolution of the cavity modes with photon annihilation and creation operators $a$ and $a^\dagger$ as well as their coupling to the external field $\ain$ is given by ($\hbar = 1$ used here and in the following)~\cite{Heeg2013b}
\begin{align}
 H_M &= \dc {a_1}^\dagger {a_1} + \dc {a_2}^\dagger {a_2} \nonumber \\ 
     &+ i \sqrt{2 \kR} \left[ (\vec{\hat a}_1^* \cdott \vec{\hat a}_\text{in}) \, a_\text{in} a_1^\dagger - (\vec{\hat a}_\text{in}^* \cdott \vec{\hat a}_1) \, a_\text{in}^* a_1 \right] \nonumber \\
     &+ i \sqrt{2 \kR} \left[ (\vec{\hat a}_2^* \cdott \vec{\hat a}_\text{in}) \, a_\text{in} a_2^\dagger - (\vec{\hat a}_\text{in}^* \cdott \vec{\hat a}_2) \, a_\text{in}^* a_2 \right] \; .\label{eq:basic_H_mode}
\end{align}
Here, the expressions in the round brackets denote scalar products of the polarization directions between the incident radiation ($\vec{\hat a}_\textrm{in}$) and the cavity modes ($\vec{\hat a}_{1}$,$\vec{\hat a}_{2}$). The mismatch between the frequencies of the cavity mode and the external field is denoted by the cavity detuning
\begin{align}
 \Delta_C(\theta) = \omega_C - \omega = \omega \left[\sqrt{\cos{(\theta)}^2 + \sin{(\theta_0)}^2} -1  \right] \;. \label{eq:cavity_detuning}
\end{align}
We emphasize that the cavity detuning can be controlled with the incidence angle $\theta$ and does not depend on the frequency of the driving field for all practical purposes. The reason for this is the special grazing incidence geometry, in which the cavity is probed, and that the nuclear resonance typically is orders of magnitude more narrow than the cavity line width.

\begin{table}[t]
 \centering
 \caption{\label{tab:layers_eit}Parameters for the two-layer cavities analyzed in this work. The geometry defined in the first column corresponds to a node--antinode configuration in which an EIT-like spectrum is expected, the second parameter set defines a antinode--node configuration in which no EIT-like spectrum was observed~\cite{Roehlsberger2012}.}
\begin{tabularx}{\columnwidth}{X c X r @{}p{1cm} X r @{}p{1cm} X}
 \hline \hline \\ [-1.5ex]
  & Material & & \multicolumn{2}{c}{Thickness [nm]} && \multicolumn{2}{c}{Thickness [nm]} & \\[0.5ex]
 \hline \\ [-1.5ex]
 &Pt& & $~~~~~~~$3 & && $~~~~~~~$3 &&\\
 &C&  & 10 &.5 && 20 & &\\
 &$^{57}$Fe& & 3  & && 3 & & \\
 &C&  & 6  &.5 && 6 &.5 &\\
 &$^{57}$Fe& & 3  & && 3 & & \\
 &C&  & 21 & && 11 &.5  &\\
 &Pt& & 10 & && 10 &  &\\[0.5ex]
 \hline \hline
\end{tabularx}
\end{table}

Furthermore, incoherent effects such as the photon loss of the cavity modes have to be included in the model. This can be done via a description in terms of the density matrix $\rho$ and Lindblad operators $\mathcal{L}[\rho]$. We define 
\begin{equation}
  \mathcal{L}[\rho, \mathcal{O}^+, \mathcal{O}^-] = \big( \mathcal{O}^+ \mathcal{O}^- \rho + \rho \mathcal{O}^+ \mathcal{O}^- -2\mathcal{O}^- \rho \mathcal{O}^+ \big) \label{eqn:lindblad_helper} \;
\end{equation}
for arbitrary operators $\mathcal{O}^+$ and $\mathcal{O}^-$. Then, the photon loss can be described via
\begin{align}
 \mathcal{L}_M[\rho] = &-\kappa\,  \mathcal{L}[\rho, a_1^\dagger, a_1] -\kappa\,  \mathcal{L}[\rho, a_2^\dagger, a_2] \label{eqn:cavity_decay} \;.
\end{align}

\subsection{Nuclei\label{sec:basic_nuclei}}
Next, we include the resonant nuclei to the description. In this work, to be specific, we refer to the commonly used M\"ossbauer isotope $^{57}$Fe with its transition at $\omega_0 = 14.4$~keV and a line width of $\gamma = 4.7$~neV. To encompass the general case, we allow for a magnetic hyperfine splitting of the nuclear resonance. Then, the ground state splits up into two states separated by the energy $\delta_g$ and four excited states with energy spacing $\delta_e$ between adjacent states. This leads to six M1 allowed transitions in the nucleus~\cite{Hannon1999}, which are summarized in Tab.~\ref{tab:transitions}. We note that the energy splitting of the ground states is in the neV range, such that both states are evenly populated at room temperature according to the Boltzmann factor $\exp{(\delta_g/k_B T)}$.

In a suitable interaction picture, the nuclear dynamics is characterized by the Hamiltonian
\begin{align}
 H_N = H_0 + H_{C} \;.
\end{align}
The diagonal part
\begin{align}
 H_0 = \sum_{n=1}^N  \bigg[ &\sum_{j=1}^2 \delta_g (j- \tfrac{3}{2}) \; \ketbra{g_j^{(n)}} \nonumber \\
     + &\sum_{j=1}^4 \left( \delta_e (j- \tfrac{5}{2}) - \Delta \right) \; \ketbra{e_j^{(n)}} \bigg] \, \label{eq:basic_H_0}
\end{align}
contains the energy detuning $\Delta = \omega - \omega_0$ and the energy splitting of the states due to the magnetic hyperfine interaction for all $N$ atoms. It is important to note that in contrast to the cavity detuning $\Delta_C$, the detuning $\Delta$ does not depend on the incidence angle $\theta$, but only on the frequency of the externally applied x-ray field. In Eq.~(\ref{eq:basic_H_0}), the index $n$ sums over all $N$ nuclei, and $j$ sums over the two ground and four excited hyperfine states, respectively. The coupling of the six transitions $\mu$ (see Tab.~\ref{tab:transitions}) to the two polarization modes are given as
\begin{align}
  H_{C} = \sum_{n=1}^N \sum_{\mu=1}^6  \Big[ ( &\vec{\hat d}_\mu^* \cdott  \vec{\hat a}_1 )\, g_\mu^{(n)} S_{\mu+}^{(n)} a_1   \nonumber \\
   + \, ( &\vec{\hat d}_\mu^* \cdott  \vec{\hat a}_2 )\, g_\mu^{(n)} S_{\mu+}^{(n)} a_2 \,+ \textrm{H.c.} \Big] \; . \label{eq:basic_H_C}
\end{align}
Here, $\vec{\hat d}_\mu$ denotes the normalized dipole moment, $S_{\mu-}^{(n)}$ [$S_{\mu+}^{(n)}$] is the nuclear lowering [raising] operator of the transition $\mu$ for atom $n$, and the coupling coefficient $g_\mu^{(n)} = g \: c_{\mu} \: e^{i\,\phi^{(n)}}$ is composed of a constant $g$, the Clebsch-Gordan coefficient $c_\mu$ of the respective transition and a phase taking into account the position of the atom $n$.

Additionally, the spontaneous decay of the excited nuclear states is taken into account via
\begin{align}
 \mathcal{L}_\text{SE}[\rho] &= \sum_{n=1}^N \mathcal{L}_\text{SE}^{(n)}[\rho] \label{eqn:spont_emission} \;, \\
 \mathcal{L}_\text{SE}^{(n)}[\rho] &= -\frac{\gamma}{2} \sum_{\mu=1}^{6} c_\mu^2 \, \mathcal{L}[\rho, S_{\mu+}^{(n)}, S_{\mu-}^{(n)}] \;.
\end{align}

\begin{table}[t]
\caption{Overview of the M1 allowed transitions in the $^{57}$Fe nucleus with transition index $\mu$. Shown are the involved states, the transition energy $\Delta E$ relative to the energy at vanishing magnetization $\omega_0$, the Clebsch-Gordan coefficient (CG) $c_\mu$ and the polarization type. Linear polarization is denoted by $\pi^0$, right (left) circular polarization as $\sigma^+$ ($\sigma^-$).\label{tab:transitions}}
\begin{tabular}{ccccc}
 \hline \hline \\[-1.5ex]
 $\quad\mu\quad$ & $\quad$Transition$\quad$ & $\quad\Delta E\quad$& $\quad$C-G$\quad$ & Polarization \\[0.5ex]
 \hline\\[-1.5ex]
 1 & $\ket{g_1}\leftrightarrow\ket{e_1}$ & $-\delta_g/2-3/2\delta_e$ & $1$ & $\sigma^-$ \\
2 & $\ket{g_1}\leftrightarrow\ket{e_2}$ & $-\delta_g/2-1/2\delta_e$ & $\sqrt{2/3}$ & $\pi^0$ \\
3 & $\ket{g_1}\leftrightarrow\ket{e_3}$ & $-\delta_g/2+1/2\delta_e$ & $\sqrt{1/3}$ & $\sigma^+$ \\
4 & $\ket{g_2}\leftrightarrow\ket{e_2}$ & $\delta_g/2-1/2\delta_e$ & $\sqrt{1/3}$ & $\sigma^-$ \\
5 & $\ket{g_2}\leftrightarrow\ket{e_3}$ & $\delta_g/2+1/2\delta_e$ & $\sqrt{2/3}$ & $\pi^0$ \\
6 & $\ket{g_2}\leftrightarrow\ket{e_4}$ & $\delta_g/2+3/2\delta_e$ & $1$ & $\sigma^+$\\[0.5ex]
\hline \hline
\end{tabular}
\end{table}

\subsection{Input-output relation and observables\label{sec:basic_io}}
In order to relate the internal operators in the cavity to externally accessible quantities, the input-output relations are employed~\cite{Gardiner2004}. The output field $\aout$, also visualized in Fig.~\ref{fig:setup}, is given by
\begin{align}
 a_\text{out} &= -a_\text{in} \, \left(\vec{\hat a}_\text{out}^* \cdott \vec{\hat a}_\text{in} \right) \nonumber \\
 &+ \sqrt{2 \kR} \left[ (\vec{\hat a}_\text{out}^* \cdott \vec{\hat a}_1) \, a_1 + (\vec{\hat a}_\text{out}^* \cdott \vec{\hat a}_2) \, a_2 \right] \,.
 \label{eq:basic_io}
\end{align}
With this operator at hand, the reflection coefficient reads
\begin{align}
 R = \frac{\ew{\aout}}{\ain} \;. \label{eq:basic_R}
\end{align}
Note that in a typical experiment, the reflectance $|R|^2$ is measured.

\subsection{Full model\label{sec:basic_full}}
The expressions given above form the building blocks of the general model developed in Ref.~\cite{Heeg2013b}. The full master equation reads
\begin{align}
 \frac{d}{dt}\rho = -i [ H_M + H_N, \rho] + \mathcal{L}_M[\rho] + \mathcal{L}_\text{SE}[\rho]\;.
 \label{eq:basic_me}
\end{align}
In principle, the dynamics of the system could be solved this way. However, due to the huge Hilbert space connected with the $N$ atoms and two cavity modes, this task is challenging. Hence, in Ref.~\cite{Heeg2013b} two approximations well justified at present experimental conditions  were performed. First, the cavity modes $a_1$ and $a_2$ were adiabatically eliminated. This is possible since the cavity modes have a low quality factor $Q$~\cite{Roehlsberger2010}, which is known as the bad-cavity regime~\cite{Meystre2007}. Hence, the time scale $1/\kappa$, on which the mode dynamics equilibrates, is very short compared to the nuclear time scale and therefore the cavity modes can be considered as stationary. Second, it is possible to restrict the analysis to the subspace of up to one excitation in the system, since experiments performed at current synchrotron radiation sources provide on average less than one resonant photon per pulse~\cite{Shenoy2008,Roehlsberger2010,Roehlsberger2013}. This simplifies the master equation considerably and compact analytic expressions for the observables can be obtained. Below we will exploit the same approaches to also simplify the extended model which will then include different resonant layers and multiple modes.

\section{Generalization to multiple layers and multiple modes\label{sec:generalization}}
We will now extend the basic model introduced in the last section by explicitly including multiple layers of resonant nuclei as well as more than a single cavity mode in theory. As already discussed above, in typical experiments at most one cavity mode can be driven resonantly at a time. This is due to their large angular separation, which significantly exceeds the beam divergence at modern x-ray sources. The angular separation in turn leads to large cavity detunings $\Delta_C$ for nonresonant modes. Nevertheless, as we will show below, the additional modes can sometimes be of importance, since the nuclei can in principle scatter into them, or if the reflectance is considered over a broad range of incidence angles. Also the inclusion of multiple layers is a highly desirable goal, as motivated by observation of EIT in Ref.~\cite{Roehlsberger2012}.

Clearly, the coefficients in the master equation, such as decay rates $\kappa$ or coupling constants $g$, differ for each mode and each layer and ought to be marked with an index for the respective element. In an attempt to reduce confusion, we stick to the following notation: An atomic index is denoted by an upper index $n$ in brackets. Lower indices $\mu$ indicate a transition, as listed in Tab.~\ref{tab:transitions}. The different cavity modes are distinguished by an upper index ${[j]}$ in squared brackets. A curly bracket $\{l\}$ indicates that the respective quantity is related to layer~$l$. This notation is summarized in Tab.~\ref{tab:notation}.

We start by revisiting the internal and external electromagnetic field. We consider a single incident field $\ain$, which impinges onto the cavity surface under the grazing angle $\theta$, and an outgoing field $\aout$, emitted at the respective reflection angle $\pi - \theta$. Compared to the initial analysis in Sec.~\ref{sec:basic}, the input field does not only drive one cavity mode $a$, but multiple modes $a^{[j]}$. At the same time, the output field is driven by these modes and, naturally, also the resonant nuclei will interact with the different cavity field modes. This means, we also have to distinguish the coupling coefficients and decay rates for each cavity mode.

\begin{table}[t]
 \centering
 \caption{\label{tab:notation}Index notation used throughout this work.}
\begin{tabularx}{\columnwidth}{X c X c X X }
 \hline \hline \\[-1.5ex]
  & Index & &  Quantity & \\[0.5ex]
 \hline \\[-1.5ex]
 & $n,m$ & & nucleus $n, m \in\{1,\dots,N\}$ & \\
 & $\mu, \nu$ & & nuclear transition $\mu, \nu \in\{1,\dots,6\}$ & \\
 & $\{l\}, \{k\}$ & & nuclear layer & \\
 & $[j]$ & & cavity mode& \\[0.5ex]
  \hline \hline
\end{tabularx}
\end{table}

Generalizing Eq.~(\ref{eq:basic_io}) from the original theory, we write for the input-output relation
\begin{align}
 \aout &= -\ain \scpr{\vec{\hat a}_\textrm{out}^*}{\vec{\hat a}_\textrm{in}} \nonumber \\  &+ \sum_j \sqrt{2\kappa_R^{[j]}} \Big[ a_1^{[j]} \scpr{\vec{\hat a}_\textrm{out}^*}{\vec{\hat a}_1^{[j]}}  + a_2^{[j]} \scpr{\vec{\hat a}_\textrm{out}^*}{\vec{\hat a}_2^{[j]}} \Big]
\end{align}
and the Hamiltonian describing the dynamics of the modes given in Eq.~(\ref{eq:basic_H_mode}) becomes
\begin{align}
 H_M = \sum_j & \Delta_C^{[j]} \left( {a_1^{[j]}}^\dagger a_1^{[j]} + {a_2^{[j]}}^\dagger a_2^{[j]} \right) + i \sum_j \sqrt{2\kappa_R^{[j]}} \nonumber \\
\times\;\Big[\;\;&
\ain {a_1^{[j]}}^{\dagger} \scpr{{{\vec{\hat a}_1^{[j]}}^*}}{\vec{\hat a}_\textrm{in}}
-\ain^*{a_1^{[j]}} \scpr{\vec{\hat a}_\textrm{in}^*}{{{\vec{\hat a}_1^{[j]}}}}
\nonumber \\ 
+\;&\ain {a_2^{[j]}}^{\dagger} \scpr{{{\vec{\hat a}_2^{[j]}}^*}}{\vec{\hat a}_\textrm{in}}
-\ain^*{a_2^{[j]}} \scpr{\vec{\hat a}_\textrm{in}^*}{{{\vec{\hat a}_2^{[j]}}}}
 \Big] \;.
\end{align}
In a similar fashion, the couplings with nuclei are modified to include the sum over all modes $j$ and the interaction Hamiltonian given in Eq.~(\ref{eq:basic_H_C}) and describing a transition $\mu$ of an atom $n$ is extended accordingly. While the coupling to the single cavity mode was denoted by $g_\mu^{(n)}$ before, the coefficients related to a general cavity mode $j$ are now named $g_\mu^{(n)[j]}$. They can be decomposed into $g_\mu^{(n)[j]} = g^{[j]} c_\mu e^{i \phi^{(n)}}$, where $c_\mu$ is the Clebsch-Gordan coefficient of the transition $\mu$, $\phi^{(n)}$ accounts for a potential phase imprinted on the nucleus by the field due to the atomic position, and $g^{[j]}$ denotes a universal coupling constant between mode $j$ and all nuclei and transitions. Note that this factorization is possible in this way only because we assumed a single thin layer of resonant nuclei in this analysis.

However, as soon as we consider multiple layers, the assumption of uniform coupling strengths $g^{[j]}$ for all atoms is clearly no longer justified. For example, for a cavity without resonant nuclei, roughly an intensity profile with $\sin^2$ shape can be expected along the cavity for the guided modes. Different layers at different positions will thus experience different field strengths and the coupling coefficient $g$ to the cavity mode cannot be considered as a constant anymore. Also, we want to emphasize that the same argument holds if a very thick layer of resonant nuclei is present in the cavity. Here, the nuclei close to the two layer boundaries might be exposed to strongly differing field strengths and the respective coupling coefficients become spatially dependent.

Both cases can be modeled by introducing several ensembles of nuclei. The atoms in each ensemble are situated at the same depth of the cavity and hence couple to the modes with a common coefficient. We denote this coupling parameter between the nuclei in the layer $l$ and the cavity modes $j$ by $g^{[j]\{l\}}$. The coupling coefficient of the transition $\mu$ in single atom $n$ located in the layer $l_n$ then reads
\begin{align}
g_{\mu}^{(n)[j]} = g^{[j]\{l_n\}} c_\mu e^{i \,\phi^{(n)}}\;,
\end{align}
The number of nuclei in each layer is $N^{\{l\}}$ and the total number of resonant nuclei is $N = \sum_l N^{\{l\}}$. In this formulation, the coupling Hamiltonian from Eq.~(\ref{eq:basic_H_C}) is generalized to
\begin{align}
H_C = \sum_{n, \mu,  j} \Big[
&\scpr{\vec{\hat d}_\mu^*}{\vec{\hat a}_1^{[j]}} g_\mu^{(n)[j]} S_{\mu+}^{(n)} a_1^{[j]} \nonumber \\ +
&\scpr{\vec{\hat d}_\mu^*}{\vec{\hat a}_2^{[j]}} g_\mu^{(n)[j]} S_{\mu+}^{(n)} a_2^{[j]}
+ \textrm{H.c.} \Big] \;.
\end{align}
The diagonal part $H_0$ containing the energy shifts of the states and the detuning $\Delta$ is unaffected by our extension of the model.

Next to the Hamiltonian dynamics, also the incoherent part capturing the mode decays need to be extended accordingly. The Lindblad operator describing the photon loss in the cavity modes, see Eq.~(\ref{eqn:cavity_decay}), becomes
\begin{align}
 \mathcal{L}_M[\rho] = &-\sum_j \kappa^{[j]}\,  \mathcal{L}[\rho, {a_1^{[j]}}^\dagger, a_1^{[j]}] \nonumber \\
 &-\sum_j \kappa^{[j]} \,  \mathcal{L}[\rho, {a_2^{[j]}}^\dagger, a_2^{[j]}] \;,
\end{align}
whereas the spontaneous emission contribution of the nuclei remains the same.

\subsection{Effective Master equation}
In a next step we simplify the master equation by applying the same approximations as in the case of the original model, which were described already in Sec.~\ref{sec:basic_full}.

First, we perform the adiabatic elimination of the cavity modes. In contrast to the basic model, we do not eliminate the two modes $a_1$ and $a_2$ for the two polarization directions only, but a total of $2 j$ modes. However, since the different modes are not directly mutually coupled, they can be eliminated independently and their contributions to the effective master equation sum up.

From the Heisenberg equation of motion for the cavity mode operators
\begin{align}
 \frac{d}{dt} a_\iota^{[j]} = i [ H_M + H_0 + H_C, a_\iota^{[j]} ] - \kappa^{[j]} a_\iota^{[j]} \;
\end{align}
we find the stationary solutions
\begin{align}
  a_\iota^{[j]} &= \frac{1}{\kappa^{[j]} + i\Delta_C^{[j]}}
 \bigg[ \sqrt{2 \kappa_R^{[j]}} \ain ({\vec{\hat a}_\iota^{[j]}}^*\cdott\vec{\hat a}_\text{in}) \nonumber \\
 &\qquad\qquad- i \sum_{n,\mu} ({\vec{\hat a}_\iota^{[j]}}^*\cdott\vec{\hat d}_\mu) {g_\mu^{(n)[j]}}^* S_{\mu-}^{(n)} \bigg] \;,
\end{align}
where $\iota = 1,2$ indicates the two perpendicular cavity mode polarizations.

Inserting these operators in the full model, we obtain the effective master equation for the nuclei
\begin{align}
 \frac{d}{dt} \rho = -i [ H_\textrm{eff}, \rho] + \mathcal{L}_\textrm{eff}[\rho] \;,
\end{align}
with the effective Hamiltonian and the Lindblad terms
\begin{align}
 H_\textrm{eff} &= H_0 + H_\Omega + H_\textrm{LS} \;, \\
 \mathcal{L}_\textrm{eff}[\rho] &= \mathcal{L}_\text{SE}[\rho] + \mathcal{L}_\text{cav}[\rho] \;.
\end{align}
In the same notation as in Ref.~\cite{Heeg2013b}, the individual components of these equations are found as
\begin{align}
 H_\Omega &= \sum_{n, \mu} \scprt{\vec{\hat d}_\mu^*}{\vec{\hat a}_\textrm{in}} \sum_j \left( \Omega^{[j]} g_\mu^{(n)[j]}\right) S_{\mu+}^{(n)} + \textrm{H.c.} \; , \label{eq:H_Omega_full} \\[2ex]
 H_\textrm{LS} &= \sum_{n,m} \sum_{\mu,\nu} \scprt{\vec{\hat d}_\mu^*}{\vec{\hat d}_\nu} \sum_j \left( \delta_\textrm{LS}^{[j]} g_\mu^{(n)[j]} {g_\nu^{(m)[j]}}^* \right) \nonumber \\ &\quad \times  S_{\mu+}^{(n)} S_{\nu-}^{(m)} \; , \\[2ex]
 \mathcal{L}_\textrm{cav}[\rho] &= \sum_{n,m} \sum_{\mu, \nu} \scprt{\vec{\hat d}_\mu^*}{\vec{\hat d}_\nu} \sum_j\left(- \zeta_{S}^{[j]} g_{\mu}^{(n)[j]} {g_{\nu}^{(m)[j]}}^{*} \right) \nonumber \\ &\quad \times \mathcal{L}[\rho, S_{\mu+}^{(n)}, S_{\nu-}^{(m)}] \;, \label{eq:L_cav_full}
\end{align}
with the coefficients
\begin{align}
 \Omega^{[j]} &= \frac{\sqrt{2 \kR^{[j]}} \ain}{\kappa^{[j]} + i\Delta_C^{[j]}} \,,\\
 \delta_\textrm{LS}^{[j]}  &= {\operatorname{Im}\left(\frac{1}{\kappa^{[j]} + i \Delta_C^{[j]}} \right)} \;, \\
 \zeta_S^{[j]} &= {\operatorname{Re}\left(\frac{1}{\kappa^{[j]} + i \Delta_C^{[j]}} \right)  }\;.
\end{align}
and the outer product $\mathbb{1}_\perp = \vec{\hat a}_1 \vec{\hat a}_1^* + \vec{\hat a}_2 \vec{\hat a}_2^*$. Note that this completeness relation only refers to the two possible mode polarizations, such that no sum over the different modes $[j]$ is required. Moreover, the adiabatic elimination also affects the input-output relation from Eq.~(\ref{eq:basic_io}) and thus the observable in Eq.~(\ref{eq:basic_R}). We obtain for the reflection coefficient
\begin{align}
R &= R_C + R_N \;,
\end{align}
with the cavity contribution $R_C$ and the nuclear part of the reflectance $R_N$. The two reflection coefficients are given by
\begin{align}
R_C &= \left(-1 + \sum_{j}\frac{2\kappa_R^{[j]}}{\kappa^{[j]} + i \Delta_C^{[j]}} \right)\scpr{\vec{\hat{a}}_\textrm{out}^* }{\vec{\hat{a}}_\textrm{in}} \;, \label{eq:R_C_multimode}\\
R_N& =  - \frac{i}{\ain} \sum_{n,\mu} \left( \sum_j\frac{\sqrt{2\kappa_R^{[j]}} {g_\mu^{(n)[j]}}^*}{\kappa^{[j]} + i \Delta_C^{[j]}} \right) \scprt{\vec{\hat{a}}_\textrm{out}^* }{\vec{\hat{d}}_\mu} \nonumber \\
&\qquad \times \langle S_{\mu-}^{(n)} \rangle \;.
\end{align}

In a second approximation, we restrict the dynamics of the system to the subspace of up to one excitation in the system. As mentioned above, the reduction to the linear regime is well justified for experiments performed at current synchrotron radiation sources. In the initial stage, all nuclei reside in one of the two hyperfine ground states $\ket{g_1}$ and $\ket{g_2}$, with equal probability at room temperature. Further, we can assume that the nuclei of the different macroscopic ensembles $l$ introduced above are evenly distributed among these two states as well. This collective ground state is denoted by $\ket G$.

The definition of the collective excited states demands a more elaborate approach. In Ref.~\cite{Heeg2013b} collective excited states $\ket{E_\mu^{(+)}}$ were introduced, which denote a symmetrized excitation on the transition $\mu$. Note that such an excitation is shared only by $N/2$ nuclei, since only half of the nuclei were originally in the ground state of the respective transition. Here, we now generalize these states and denote a collectively excited state in the ensemble $l$ on the transition $\mu$ by $\ket{E_\mu^{\{l\}}}$. More formally, we define it as
\begin{align}
 \ket{E_\mu^{\{l\}}} = \frac{1}{\sqrt{N^{\{l\}}/2}} \: \sum_{n}^{N^{\{l\}}/2}  e^{i\,\phi^{(n)}} \: S_{\mu + }^{(n)} \: \ket{G} \;,
\end{align}
and again only half of the nuclei in the respective layer $l$ contribute due to the ground state distribution. Each of the contributory nuclei couples to the modes with the same rate $g^{[j]\{l\}}$. This qualifies the collective states defined here to rewrite the system dynamics given in Eqs.~(\ref{eq:H_Omega_full})--(\ref{eq:L_cav_full}) in the linear regime. In the process, the sums over the macroscopic number of atoms $\sum_n$ is simplified to the much more manageable sum over the different resonant layers $l$ as
\begin{align}
 \sum_n g_{\mu}^{(n)[j]} S_{\mu+}^{(n)} = \sum_l \sqrt{\tfrac{1}{2} N^{\{l\}}} c_{\mu} g^{[j]\{l\}} \ket{E_\mu^{\{l\}}}\bra{G} \;.
\end{align}
This way, we obtain the effective equations for the linear regime
\begin{align}
H_\Omega &= \sum_{\mu, j, l} \scprt{\vec{\hat d}_\mu^*}{\vec{\hat a}_\textrm{in}} \left( \Omega^{[j]} c_{\mu} g^{[j]\{l\}}\right) \nonumber \\
&\quad \times \sqrt{\tfrac{1}{2}N^{\{l\}}}\: \ket{E_\mu^{\{l\}}}\bra{G}+ \textrm{H.c.} \;, \label{eq:H_Omega_lin} \\
H_\textrm{LS} &= \sum_{\mu,\nu} \sum_{j,l,k} \scprt{\vec{\hat d}_\mu^*}{\vec{\hat d}_\nu} \left( \delta_\textrm{LS}^{[j]} c_{\mu}c_{\nu} g^{[j]\{l\}} {g^{[j]\{k\}}}^* \right) \nonumber \\
&\quad \times \tfrac{1}{2}\sqrt{ N^{\{l\}} N^{\{k\}} }\: \ket{E_\mu^{\{l\}}}\bra{E_\nu^{\{k\}}} \; ,  \\[2ex]
\mathcal{L}_\textrm{cav}[\rho] &= \sum_{\mu, \nu} \sum_{j,l,k} \scprt{\vec{\hat d}_\mu^*}{\vec{\hat d}_\nu} \left(-\zeta_{S}^{[j]} c_{\mu}c_{\nu} g^{[j]\{l\}} {g^{[j]\{k\}}}^{*} \right) \nonumber \\
&\quad \times \tfrac{1}{2}\sqrt{ N^{\{l\}} N^{\{k\}} }\: \mathcal{L}[\rho, \ket{E_\mu^{\{l\}}}\bra{G}, \ket{G}\bra{E_\nu^{\{k\}}}] \;. \label{eq:Lcav_lin}
\end{align}
Finally, the reflection coefficient reads
\begin{align}
  R = R_C &-\frac{i}{\ain} \sum_{\mu,j,l} \left( \frac{\sqrt{2\kappa_R^{[j]}} c_{\mu} {g^{[j]\{l\}}}^*}{\kappa^{[j]} + i \Delta_C^{[j]}} \right) 
\nonumber \\ &\quad \times
\scprt{\vec{\hat{a}}_\textrm{out}^* }{\vec{\hat{d}}_\mu} \sqrt{\tfrac{1}{2}N^{\{l\}}}\: \bra{E_\mu^{\{l\}}}\rho\ket{G} \;. \label{eq:R_lin}
\end{align}

\subsection{Unmagnetized layers}
A commonly encountered scenario is the setting without a magnetic hyperfine splitting in the resonant layers. In this case, the level scheme of the $^{57}$Fe nucleus reduces to a two-level system with one ground and one excited state. From the general theory above, this behavior can be emulated by setting the energy splittings $\delta_g$, $\delta_e$ of the ground and excited states to zero and choosing the quantization axis such that the incident beam with polarization $\vec{\hat a}_\textrm{in}$ only drives the linearly polarized transitions $\mu = 2,5$ (c.f.~Tab.~\ref{tab:transitions}). Further, we define the state
\begin{align}
 \ket{E^{\{l\}}} = \frac{1}{\sqrt{2}} \left( \ket{E_2^{\{l\}}} + \ket{E_5^{\{l\}}} \right) \;,
\end{align}
which describes an excitation in the $l$th resonant layer without the distinction of the two hyperfine substates. We obtain for the master equation and the reflection coefficient
\begin{align}
  H_\Omega &= \sum_j \Omega^{[j]}\sqrt{\tfrac{2}{3}} \sum_l g^{[j]\{l\}} \sqrt{N^{\{l\}}} \ket{E^{\{l\}}}\bra{G} + \textrm{H.c.} \;, \label{eq:H_Omega_lin_B0} \\[2ex]
 H_\textrm{LS} &= \sum_j \delta_\textrm{LS}^{[j]} \tfrac{2}{3} \sum_{l,k} g^{[j]\{l\}} {g^{[j]\{k\}}}^* \sqrt{N^{\{l\}} N^{\{k\}}} \nonumber \\ &\quad\quad\times \ket{E^{\{l\}}}\bra{E^{\{k\}}}  \;, \label{eq:H_LS_lin_B0} \\[2ex]
 \mathcal{L}_\textrm{cav}[\rho] &= -\sum_j \zeta_{S}^{[j]} \tfrac{2}{3} \sum_{l,k} g^{[j]\{l\}} {g^{[j]\{k\}}}^* \sqrt{N^{\{l\}} N^{\{k\}}} \nonumber \\ &\quad\quad\times \mathcal{L}[\rho, \ket{E^{\{l\}}}\bra{G},\ket{G}\bra{E^{\{k\}}}] \label{eq:Lcav_lin_B0} \;, \\[2ex]
 R &= \bigg[ -1 + \sum_j\frac{2\kappa_R^{[j]}}{\kappa^{[j]} + i\Delta_C^{[j]}}  - \frac{i}{\ain} \sum_j \frac{\sqrt{2\kappa_R^{[j]}}}{\kappa^{[j]}+i\Delta_C^{[j]}} \nonumber \\ &\times \sqrt{\tfrac{2}{3}} \sum_l {g^{[j]\{l\}}}^* \sqrt{N^{\{l\}}} \bra{E^{\{l\}}}\rho\ket{G} \bigg] \scpr{\vec{\hat{a}}_\textrm{out}^* }{\vec{\hat{a}}_\textrm{in}} \;. \label{eq:R_lin_B0}
\end{align}
This set of equations will form the basis for the description of the EIT experiment~\cite{Roehlsberger2012}, which we will analyze in Sec.~\ref{sec:eit}.

\subsection{Heuristic extensions\label{sec:heuristic}}

Before we study the phenomena and consequences which emerge from our generalized theory, we take a step back and consider additional effects in our system, which in general will turn out to be of significance. We emphasize, that so far our theory is developed to capture the cavity character of the layer system, i.e.~the guided modes and the embedded resonant nuclei. However, in the grazing incidence geometry, also effects which are not related to the structure of the cavity and stem from bulk material properties become important.

Since the refractive index of the cavity materials at x-ray energies is less than one, total reflection is observed for small incidence angles $\theta$ in the few-mrad range, while for larger angles the light is essentially completely absorbed. The transition between those two regimes is not sudden, but can be characterized by a smooth function $R_\textrm{Envelope}(\theta)$. As soon as we consider the reflectance over a broader range of incidence angles, this envelope has to be taken into account. Note that previous studies have been performed at a fixed incidence angle~\cite{Roehlsberger2010,Roehlsberger2012,Heeg2013,Heeg2014} or covered only tiny angular ranges, for which the envelope could be considered constant~\cite{Heeg2014b}. To include this total reflection behavior at grazing incidence,
we will heuristically combine the analytical formula from Eq.~(\ref{eq:R_C_multimode}), describing the guided modes, with the reflection curve of the cavity's mirror material only, which approximately takes into account the total reflection envelope. The envelope function $R_\textrm{Envelope}(\theta)$ is described in more detail in Appendix~\ref{sec:appendix_envelope}.

Furthermore, it has been observed that the cavity material dispersion leads to an additional relative phase between light reflected off of the outside of the cavity and light entering the cavity. This dispersion phase was found to be necessary, e.g., to describe the asymmetry of the reflection curve $R(\theta)$ around the minima of the guided modes~\cite{dispersionphase}. It can be included by generalizing the contribution which stems from the direct reflection on the cavity surface (``$-1$'' in Eq.~(\ref{eq:R_C_multimode})) with an additional phase factor $\exp{(i\phi_C)}$. As a second heuristic extension, we include such a phase phase factor as well, and allow for a complex variable $r$ instead of the cavity surface amplitude $-1$ with $|r|\approx 1$. This has the additional advantage that such a modification can also take into account possible effects of far off-resonant modes, which would give rise to a small constant offset to the reflection coefficient.

With the heuristic modifications described above, the cavity contribution to the reflection coefficients reads
\begin{align}
 R_C(\theta) = R_\textrm{Envelope}(\theta) \, \left( r + \sum_j\frac{2\kappa_R^{[j]}}{\kappa^{[j]} + i\Delta_C^{[j]}(\theta)} \right)  \;, \label{eq:R_C_multimode_heuristic}
\end{align}
with the cavity detuning (c.f.~Eq.~(\ref{eq:cavity_detuning}))
\begin{align}
 \Delta_C^{[j]}(\theta) = \omega \left[\sqrt{\cos{(\theta)}^2 + \sin{(\theta_0^{[j]})}^2} -1  \right] \;.
\end{align}

\section{Effect of multiple modes\label{sec:effect_modes}}
In this part, we will discuss the influence of multiple modes on the reflectivity, while we still restrict ourselves to a single thin layer of resonant nuclei in the cavity. Hence, the index $l$ corresponding to the different nuclear ensembles in the cavity, will be omitted in the following.

We start by considering the nuclear contribution to the reflectance only. Restricting ourselves to only one layer $l$, it can be easily seen from Eqs.~(\ref{eq:H_Omega_lin})--(\ref{eq:R_lin}) that the general form of contributions to the effective master equation does not depend on the number of cavity modes $j$. In particular, no new operators or additional couplings between the different collective states are present in the effective master equation, which is a direct consequence of the adiabatic elimination. The sole differences are the coefficients entering the expressions. For instance, generalizing the driving Hamiltonian $H_\Omega$ from a single to multiple modes requires only the modification
\begin{align}
 \Omega g_\mu^{(n)} \;\rightarrow\; \sum_j \Omega^{[j]} g_\mu^{[j]} 
\end{align}
on the level of a coefficient. Similar replacements of the variables are required for the other parts contributing to the master equation.

With the knowledge that the basic equations in the cases with one and with multiple cavity modes $j$ are equivalent, the results obtained in the original theory (c.f.~Ref.~\cite{Heeg2013b}) can be straightforwardly extended by replacing the coefficients with their respective generalized counterparts. Doing so for the linear reflectance without magnetic hyperfine splitting and neglecting the trivial polarization dependency $\scpr{\vec{\hat{a}}_\textrm{out}^* }{\vec{\hat{a}}_\textrm{in}}$, this yields
\begin{align}
 R_N& = -i \frac{2N}{3} \frac{
\left( \sum_j \tfrac{\sqrt{2\kappa_R^{[j]}}g^{[j]}}{\kappa^{[j]}+i\Delta_C^{[j]}} \right)\left(
\sum_j \tfrac{\sqrt{2\kappa_R^{[j]}}{g^{[j]}}^{*}}{\kappa^{[j]}+i\Delta_C^{[j]}}\right)
}{\Delta + i\tfrac{\gamma}{2} + \tfrac{2 N}{3} \left[ \sum_j \left|g^{[j]}\right|^2 \left( i \zeta_{S}^{[j]} - \delta_\textrm{LS}^{[j]} \right) \right]} \;.
\label{eq:multimode_RN}
\end{align}

Since in the single-mode theory a Lorentzian profile was derived for the nuclear spectrum, we also recover this line shape here in the case of multiple cavity modes. Similar to the original model, it is shifted due to a collective Lamb shift and broadened due to superradiance~\cite{Roehlsberger2010}. It can be seen from the denominator in Eq.~(\ref{eq:multimode_RN}) that each mode induces its own frequency shift and line broadening. But typically for any angle of incidence, all but (at most) one mode are driven far off-resonantly as mentioned above. Then, the according values for the cavity detuning $\Delta_C$ become large and their respective contributions to the cooperative Lamb shift and to the superradiance diminish. From the numerator of the nuclear part, we find that the strength of the nuclear signal is typically determined by one dominant mode with the smallest $\Delta_C$. Generally, the behavior is as follows. The nuclei can be excited by the external driving field via each mode $j$. This is represented by the first sum in the numerator of Eq.~(\ref{eq:multimode_RN}). The emission forms an independent second step and can again occur via each mode, as indicated by the second sum. Hence, in the general case, interferences between the different cavity modes can arise. Nevertheless, the general Lorentzian structure of the line profile is unaffected by this and hence no qualitatively different features appear in the spectrum.

The main difference to the single-mode result is found in the cavity contribution to the reflectance $R_C$ given in Eq.~(\ref{eq:R_C_multimode}), and accordingly its heuristic extension in Eq.~(\ref{eq:R_C_multimode_heuristic}). Since we included multiple guided modes in the analysis above, it is clear that the resonances of these modes should become apparent in the reflection curve, i.e.~when considering the cavity reflectance in dependence on the incidence angle $\theta$. Indeed, the expressions we derived in Eqs.~(\ref{eq:R_C_multimode}) and~(\ref{eq:R_C_multimode_heuristic}) highlight these resonances in its sum. The resonance of a guided mode $j$ is encountered at $\theta = \theta_0^{[j]}$, where $\Delta_C^{[j]} = 0$, and the reflection curve will exhibit a local minimum in the vicinity of the resonant angles.

At this point it is of interest, how well the actual angular dependent reflection curve $R(\theta)$ can be described by the cavity part of the reflectance. To this end, we numerically calculate the reflection curve using established semiclassical methods, such as Parratt's formalism~\cite{Parratt1954} or simulations by \textsc{conuss}~\cite{Sturhahn2000}, which both give equivalent results.
As we will describe the EIT scenario from Ref.~\cite{Roehlsberger2012} below,  we specialize to this particular cavity structure. The parameters of this cavity geometry are given in the middle column of Tab.~\ref{tab:layers_eit}. Note that the two resonant iron layers do not pose a challenge in this analysis, since in this first step the nuclear resonances are omitted in the description of the angular dependent reflection curve.

\begin{figure}[t]
 \centering
 \includegraphics[scale=1]{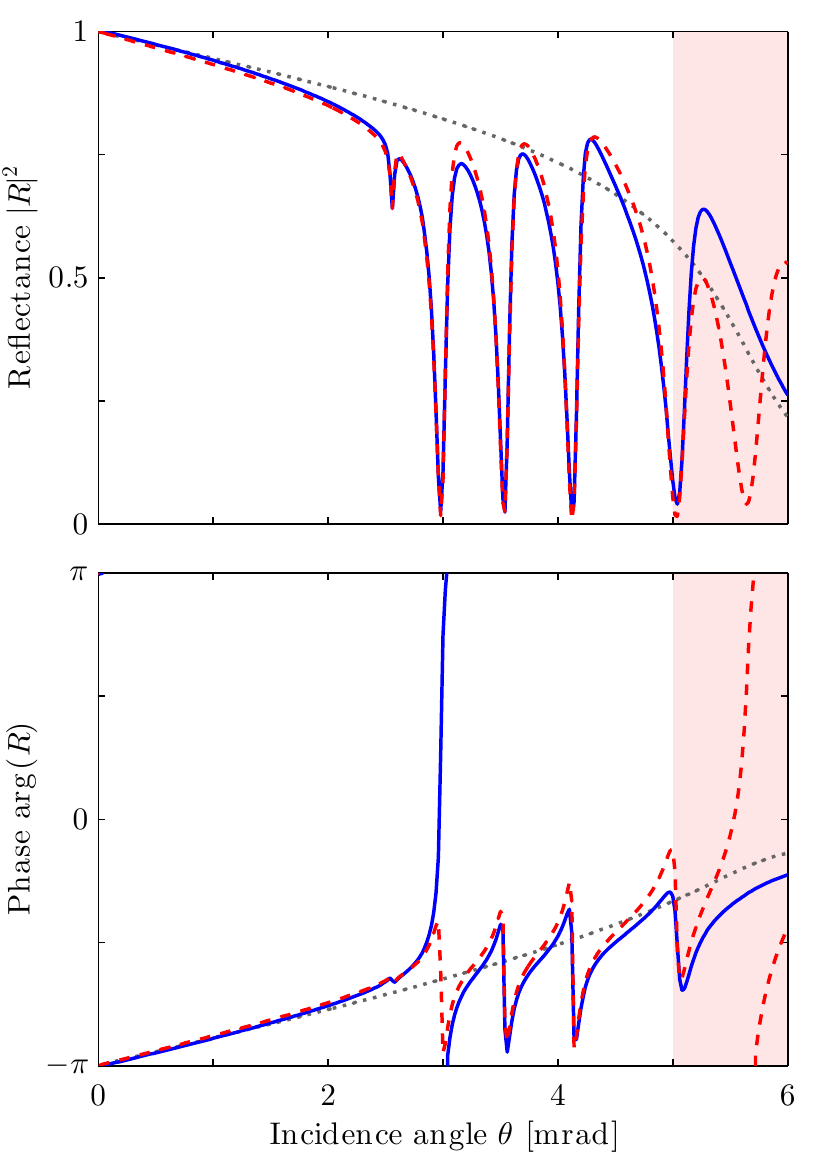}
 \caption{\label{fig:eit_rocking_qo_parratt}(Color online) Cavity reflectance as a function of the incidence angle $\theta$. The minima denote resonance angles at which guided modes are driven resonantly. Including a dispersion phase and the heuristic extension of an envelope given by the topmost layer (gray dotted line), the quantum optical theory (blue, solid) can reproduce the exact result calculated with Parratt's formalism (red, dashed) very well. In the quantum optical model only the first five guided modes were taken into account and the fit was restricted to the range $0 \le \theta \le 5$~mrad, as indicated by the shaded area. Although only the absolute value of the reflectance was fitted, the behavior of the phases calculated with the two descriptions are very similar.}
\end{figure}

In the specified cavity it is platinum that acts as cavity mirror material, therefore the envelope function $R_\textrm{Envelope}(\theta)$ taking account for the total reflection is the reflectivity of a single infinitely thick Pt layer. We fitted the absolute value of Eq.~(\ref{eq:R_C_multimode_heuristic}) with a maximum number of cavity modes $j=5$ to the expected reflection curve for the cavity, which was obtained using Parratt's formalism. The fit was performed in the range $0 \le \theta \le 5$~mrad and the obtained parameters are summarized in Appendix~\ref{sec:appendix_parameters}. The result is shown in Fig.~\ref{fig:eit_rocking_qo_parratt}. Clearly, the quantum optical model together with the heuristic extensions are well suited to describe the reflection curve. We observe that the last guided mode in Fig.~\ref{fig:eit_rocking_qo_parratt} is not reproduced. However, this is expected, since it is the sixth mode not included in the fit model, and since its angular position is not within the fit range.

Interestingly, we find that also the phase behavior of the cavity reflection coefficient is reproduced very well. This is remarkable, because only absolute values were taken into account in the fit procedure. The analytic formula in Eq.~(\ref{eq:R_C_multimode_heuristic}) has only been corrected for a global phase to match the phase behavior predicted by the Parratt formalism. Considering the phases in more detail, a deviation can be seen at the second guided mode at $\theta \approx 3$~mrad. In contrast to Parratt's formalism, the curve obtained with the quantum optical model features an apparent phase jump of $2\pi$ at the resonance.
To understand this artifact we note that in the vicinity around a guided mode, the reflection coefficient can be approximated as $R_C \approx -1 + 2\kappa_R / [\kappa + i \Delta_C]$, which results in a minimum at the resonance angle where $\Delta_C = 0$. The reflectance vanishes completely for $2\kappa_R = \kappa$, which is known as the critical coupling condition. However, a residual reflectance occurs for both $2 \kappa_R > \kappa$ and $2 \kappa_R < \kappa$, which correspond to the over- and undercritically coupled cases, respectively~\cite{Dayan2008}. Looking solely at the modulus $|R_C|$, though, these two cases cannot be distinguished. The difference becomes only apparent when the phase of $R_C$ around the resonance angle is considered: For an undercritically coupled cavity mode the phase remains in the same branch, however it undergoes an evolution to the next branch in the overcritically coupled case, which manifests as an apparent phase jump of $2\pi$. Generally, it might be beneficial to not fit absolute values, but to use the complex values of the reflection curve instead. In this case also the over- and undercritically coupled modes should be captured correctly within the quantum optical description. However, since we are interested mainly at the third guided mode later on, which is the mode at which the EIT spectra have been measured in Ref.~\cite{Roehlsberger2012}, we will use the parameters obtained in the fit discussed above for our further analysis.

Finally, it should be mentioned that it is not meaningful to extend the quantum optical descriptions to very large incident angles $\theta$. On the one hand, the theoretical description of the perpendicular polarization directions might break down. On the other hand, distinct non-grazing incidence effects are expected, since the cavity is no longer probed in (000) Bragg geometry. The angular cutoff should therefore be around the total reflection edge, which for typical cavity settings limits the number of guided modes to approximately five.

\section{Effect of multiple resonant layers\label{sec:effect_layers}}

In this part we will examine the influence of multiple layers of resonant nuclei in the cavity, while restricting the analysis to only on cavity mode. Hence, we drop the index $j$ in the coefficients throughout this section.

Since the cavity reflection part $R_C$ is unaffected by including multiple layers in the theory, we considering the nuclear contribution to the reflectance only. Starting from the general expressions given in Eqs.~(\ref{eq:H_Omega_lin})--(\ref{eq:R_lin}) and taking only one cavity mode $j$ into account, we observe that the set of equations can be considerably simplified by means of a basis transformation.

We introduce the states which resemble a collective excitation on transition $\mu$, which is distributed among the different nuclear ensembles $l$, as
\begin{align}
 \ket{E_\mu^{\{+\}}} = \frac{1}{{\mathcal{G}}} \sum_l g^{\{l\}} \sqrt{N^{\{l\}}} \ket{E_\mu^{\{l\}}} \; \label{eq:multilayer_coll_layer_state}
\end{align}
with the normalization factor
\begin{align}
 \mathcal{G} = \Big({\sum_l \big|g^{\{l\}}\big|^2 N^{\{l\}}}\Big)^{1/2} \;.
\end{align}
Using the new states, Eqs.~(\ref{eq:H_Omega_lin})--(\ref{eq:R_lin}) become
\begin{align}
 H_\Omega &= \sqrt{\tfrac{1}{2}} \Omega\mathcal{G} \sum_\mu \scprt{\vec{\hat d}_\mu^*}{\vec{\hat a}_\textrm{in}} c_\mu  \: \ket{E_\mu^{\{+\}}}\bra{G}+ \textrm{H.c.}  \;,\\[2ex]
 H_\textrm{LS} &= \tfrac{1}{2} \delta_\textrm{LS} \mathcal{G}^2  \sum_{\mu,\nu}  \scprt{\vec{\hat d}_\mu^*}{\vec{\hat d}_\nu} c_\mu c_\nu \: \ket{E_\mu^{\{+\}}}\bra{E_\nu^{\{+\}}} \; , \\[2ex]
 \mathcal{L}_\textrm{cav}[\rho] &= -\tfrac{1}{2} \zeta_{S} \mathcal{G}^2 \sum_{\mu,\nu} \scprt{\vec{\hat d}_\mu^*}{\vec{\hat d}_\nu} c_\mu c_\nu \nonumber \\
&\qquad \times \mathcal{L}[\rho, \ket{E_\mu^{\{+\}}}\bra{G}, \ket{G}\bra{E_\nu^{\{+\}}}] \;, \\[2ex]
 R &= R_C - \frac{i}{\ain} \frac{\sqrt{\kappa_R}}{\kappa+i\Delta_C} \mathcal{G} \sum_\mu c_\mu \nonumber \\
&\qquad \times
\scprt{\vec{\hat{a}}_\textrm{out}^* }{\vec{\hat{d}}_\mu} \: \bra{E_\mu^{\{+\}}}\rho\ket{G}  \;.
\end{align}
Comparing these expressions with the effective master equation of the original single-layer theory (c.f.~Ref.~\cite{Heeg2013b}), we observe an exact correspondence of the structure. Similar as in the case of the extension to multiple cavity modes, the differences manifest only in terms of the coefficients. In particular, the collective coupling between the cavity mode and the nuclei is modified as
\begin{align}
|g|\sqrt{N} \rightarrow \mathcal{G} = \Big(\sum_l \big|g^{\{l\}}\big|^2 N^{\{l\}}\Big)^{1/2} \;,
\end{align}
while all other relations remain the same. Hence, the shape of the reflection coefficient is unaffected by taking into account multiple resonant layers. In the absence of magnetization, it is given by
\begin{align}
R& = \bigg[  -1 + \frac{2\kappa_R}{\kappa + i\Delta_C} \nonumber \\ &-i\frac{2\kappa_R}{(\kappa + i\Delta_C)^2} \frac{\tfrac{2}{3} \mathcal{G}^2 }{\Delta + i\tfrac{\gamma}{2} + \tfrac{2}{3} \mathcal{G}^2 ( i\zeta_{S} - \delta_\textrm{LS}) }  \bigg] \scpr{\vec{\hat{a}}_\textrm{out}^* }{\vec{\hat{a}}_\textrm{in}} \;. \label{eq:multilayer_R}
\end{align}
Restricting ourselves to only one layer, the coefficient $\mathcal{G}^2$ reduces to $|g|^2 N$ and we recover the result which we already derived in Ref.~\cite{Heeg2013b}.

Even though we included multiple layers in our analysis, we see from Eq.~(\ref{eq:multilayer_R}) that it is not possible to explain an EIT-like spectrum as reported in Ref.~\cite{Roehlsberger2012}. Rather, we will find that it is the combined extension of multiple layers and multiple guided modes to the theory, which will be able to explain the EIT phenomenon. This will be shown in the following Sections.

\section{Effect of multiple resonant layers and multiple modes\label{sec:effect_both}}

In the last Sections we generalized the theoretical description to include multiple modes and multiple resonant layers, respectively. When restricting to one extension at a time, we observed that both give rise to additions in the nuclear reflection amplitude, while, however, leaving the general structure of a Lorentzian line shape unaffected, c.f.~Eqs.~(\ref{eq:multimode_RN}) and (\ref{eq:multilayer_R}).

The general expressions for covering both multiple modes and multiple layers in the cavity were given in Eqs.~(\ref{eq:H_Omega_lin})--(\ref{eq:R_lin}), or, in the absence of magnetization, in Eqs.~(\ref{eq:H_Omega_lin_B0})--(\ref{eq:R_lin_B0}). To simplify this set of equations, it would be desirable to perform a basis transformation which converts the different states $\ket{E_\mu^{\{l\}}}$ which describes an excitation in a single layer into a collective layer state, similar to the transformation we performed in Eq.~(\ref{eq:multilayer_coll_layer_state}). For that purpose, one would have to sum over the layers $l$, which then contains the coupling factor $g^{[j]\{l\}}$. But since this coupling coefficient now also depends on the guided mode index~$j$, the basis transformation must also involve the sum over the modes $\sum_j$. However, it can be easily seen from Eqs.~(\ref{eq:H_Omega_lin_B0})--(\ref{eq:R_lin_B0}) that this sum would be different for every contribution to the equations of motion, since the prefactors depending on $j$ are mutually different. Hence, also in the absence of magnetization it is not possible to transform the system into a form in which only one collective state is excited. Rather, in a cavity configuration with $l$ resonant layers the equations of motion need to be solved for the $l$ coupled states $\ket{E^{\{l\}}}$. This implies that the response of the nuclear ensemble will generally go beyond a Lorentzian line profile.

The different coupling coefficients ${g^{[j]\{l\}}}$ required for the extended theory need to be determined in different ways. Apart from a direct fit to numerical data, a more sophisticated approach is to derive the relative weights and phases from the field amplitudes calculated with Parratt's formalism~\cite{deBoer1991}. This self-consistent method will be applied and explained in more detail in Sec.~\ref{sec:numerical_analysis}.

\section{Application to the EIT setting\label{sec:eit}}
Next, we will analyze a particular setting, in which both multiple layers and multiple modes are considered. In the previous Section, we already discovered that the general theory developed here allows for reflection spectra, which comprise nuclear responses beyond a simple Lorentz profile. As it was shown in Ref.~\cite{Roehlsberger2012} employing a semiclassical model, the reflectance of a cavity with two unmagnetized resonant layers can even exhibit EIT-like spectra. It is this particular scenario which will be discussed in more detail in the following.

\subsection{The EIT experiment}

\begin{figure*}[t]
 \centering
 \includegraphics[scale=0.5]{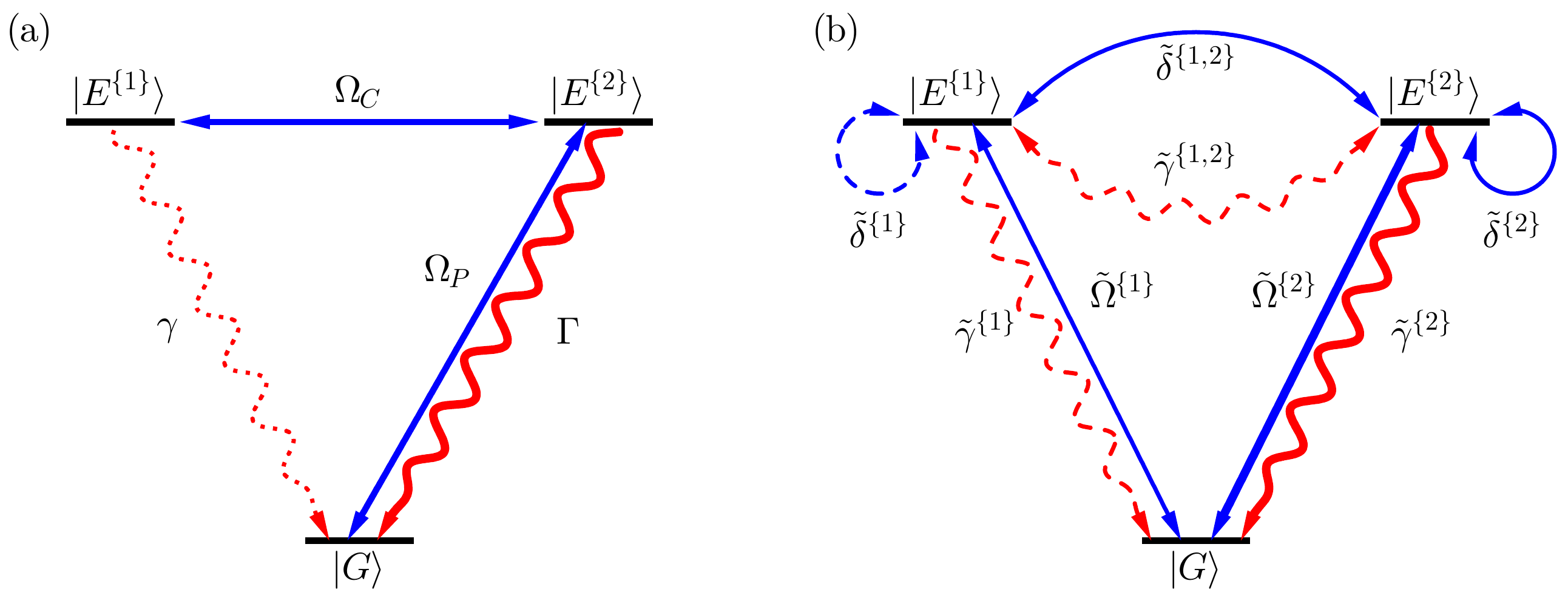}
 \caption{\label{fig:level_schemes}(Color online) (a) Effective level scheme of the EIT scenario from Ref.~\cite{Roehlsberger2012}. The nuclei in the field antinode are driven by the external field $\Omega_P$ and decay superradiantly with rate $\Gamma$, the nuclei in the field only due to spontaneous decay $\gamma$. Both ensembles are coupled via the cavity field $\Omega_C$, which gives rise to a scheme equivalent to EIT. (b) Effective level scheme in the quantum optical model. Two collective excited states are coupled to the ground state. Coherent driving and collective Lamb shifts are marked in blue, superradiant spontaneous emission is denoted by curly red single headed arrows, cross-damping between the excited states by curly red double headed arrows. [Thick solid~/ solid~/ dashed] lines denote the cavity mode detuning scalings $\sim$ [$1$~/ $\Delta_C^{-1}$~/ $\Delta_C^{-2}$] and mark the relative magnitude of the different couplings.}
\end{figure*}

The key of the setting studied in Ref.~\cite{Roehlsberger2012} is the placement of two ensembles of $^{57}$Fe nuclei in the cavity. The incidence angle was chosen such that the third guided mode of the cavity is driven resonantly and the first layer of the resonant nuclei was placed in a field node, the second layer in a antinode as sketched in Fig.~\ref{fig:field}(a). Following the interpretation in the same reference, only the latter ensemble is probed by the x-ray beam and decays superradiantly with rate $\Gamma$, while the nuclei in the first layer are only subjected to natural decay $\gamma$ on a much longer timescale. However, the first ensemble can crucially influence the system's dynamics, as a control field $\Omega_C$ between the two layers is naturally established. It arises due to radiative coupling between the two ensembles. The resulting level scheme, as also visualized in Fig.~\ref{fig:level_schemes}(a), is equivalent to a system featuring EIT. By employing a cavity like the one sketched in Fig.~\ref{fig:field}(a), the key signature of EIT, transparency of the medium on resonance, could be verified in Ref.~\cite{Roehlsberger2012}. Interestingly, it was found that the coupling field $\Omega_C$ vanishes by interchanging the roles of the two layers, i.e.~placing the first layer in a field antinode and the second ensemble in a field node, see Fig.~\ref{fig:field}(b).

\subsection{Theoretical analysis}

In order to describe the setting of two resonant layers with our quantum optical model, we restrict ourselves to two ensembles of nuclei $l=1,2$ in the cavity, but we still allow for an arbitrary number of cavity modes $j$. As before, we consider the linear response case without magnetization and omit the trivial polarization dependence in the following. We rewrite the effective Hamiltonian from Eqs.~(\ref{eq:H_Omega_lin_B0}) and~(\ref{eq:H_LS_lin_B0}) as well as the detuning part from Eq.~(\ref{eq:basic_H_0}) as
\begin{align}
 H &= \left(\tilde{\Omega}^{\{1\}} \ket{E^{\{1\}}}\bra{G} + h.c. \right)
 + \left(\tilde{\Omega}^{\{2\}} \ket{E^{\{2\}}}\bra{G} + h.c. \right) \nonumber \\
 &+  (\tilde{\delta}^{\{1\}} - \Delta) \ketbra{E^{\{1\}}} 
 +  (\tilde{\delta}^{\{2\}} - \Delta) \ketbra{E^{\{2\}}} \nonumber \\
 &+ \left(\tilde{\delta}^{\{1,2\}} \ket{E^{\{1\}}}\bra{E^{\{2\}}} + h.c. \right) \;. \label{eq:multi_eit_H}
\end{align}
Here, the first line covers the driving of the two layers, the second line accounts for the cooperative Lamb shifts and the detuning, and the last line describes a coherent coupling between the two layers. Later, we will see that the last contribution can in parts be identified with the control field $\Omega_C$ from the EIT interpretation in Ref.~\cite{Roehlsberger2012}. The incoherent Lindblad terms in our description are given by
\begin{alignat}{3}
 \mathcal{L} = -\Big(\frac{\gamma}{2} &+ \tilde{\gamma}^{\{1\}} \Big) &&\mathcal{L}[\rho, \ket{E^{\{1\}}}\bra{G},\ket{G}\bra{E^{\{1\}}}] \nonumber \\
-\Big(\frac{\gamma}{2} &+ \tilde{\gamma}^{\{2\}} \Big) &&\mathcal{L}[\rho, \ket{E^{\{2\}}}\bra{G},\ket{G}\bra{E^{\{2\}}}] \nonumber \\
&- \tilde{\gamma}^{\{1,2\}} &&\mathcal{L}[\rho, \ket{E^{\{1\}}}\bra{G},\ket{G}\bra{E^{\{2\}}}]  \nonumber \\
 &-{{}\tilde{\gamma}^{{\{1,2\}}}}^* &&\mathcal{L}[\rho, \ket{E^{\{2\}}}\bra{G},\ket{G}\bra{E^{\{1\}}}] \;. \label{eq:multi_eit_L}
\end{alignat}
Here, the first line accounts for spontaneous emission and superradiance. The other two terms describe an incoherent cross-damping term~\cite{Kiffner2010,Heeg2013}, which will contribute to the control field coupling in the EIT interpretation as well. The coefficients in Eqs.~(\ref{eq:multi_eit_H}) and~(\ref{eq:multi_eit_L}) are given by
\begin{align}
  \tilde{\Omega}^{\{l\}} &= \sum_j \Omega^{[j]}\sqrt{\tfrac{2}{3}} g^{[j]\{l\}} \sqrt{N^{\{l\}}} \;,\label{eq:multi_eit_omega_def}\\
  \tilde{\delta}^{\{l\}} &= \sum_j \delta_\textrm{LS}^{[j]} \tfrac{2}{3} \big|g^{[j]\{l\}}\big|^2 N^{\{l\}} \;,\\
  \tilde{\delta}^{\{1,2\}} &= \sum_j \delta_\textrm{LS}^{[j]} \tfrac{2}{3} g^{[j]\{1\}} {g^{[j]\{2\}}}^* \sqrt{N^{\{1\}} N^{\{2\}}} \;, \label{eq:multi_eit_d12_def} \\ 
  \tilde{\gamma}^{\{l\}} &= \sum_j \zeta_{S}^{[j]} \tfrac{2}{3} \big| g^{[j]\{l\}}\big|^2 N^{\{l\}}  \;,\\
  \tilde{\gamma}^{\{1,2\}} &= \sum_j \zeta_{S}^{[j]} \tfrac{2}{3} g^{[j]\{1\}} {g^{[j]\{2\}}}^* \sqrt{N^{\{1\}} N^{\{2\}}} \;.\label{eq:multi_eit_g12_def}
\end{align}
The effective level scheme of the system defined above is visualized in Fig.~\ref{fig:level_schemes}(b). The similarity to the scheme used in the interpretation of Ref.~\cite{Roehlsberger2012} can already be anticipated. However, in our approach a larger number of coherent and incoherent couplings are present. Nevertheless, the relative strength and hence the importance of the coupling rates can be straightforwardly estimated, as we will show in the following.

As mentioned above, in the cavity geometries of interest, the $^{57}$Fe layers are arranged such that one layer $l=1$ is located at a field node of the third guided mode, while a second layer $l=2$ is located at an antinode.
As a consequence, the nuclei in the node hardly couple to driven mode. In our quantum optical language, we can represent this idealized case by setting the respective coupling constant to zero, i.e.~$g^{[3]\{1\}} = 0$. At the same time, all other modes $j\neq 3$ are driven strongly off-resonant, such that their cavity detuning $\Delta_C^{[j]}$ becomes large. Indicating this suppression due to the large detuning with a symbolic notation $1/\Delta_C$, we find the scalings
\begin{align}
 \tilde{\Omega}^{\{1\}} \;,\; \tilde{\delta}^{\{1\}}\;,\; \tilde{\delta}^{\{1,2\}} &\sim \frac{1}{\Delta_C}  \;,\\
 \tilde{\gamma}^{\{1\}}  \;,\; \tilde{\gamma}^{\{1,2\}} &\sim \frac{1}{\Delta_C^2} \;. 
\end{align}
In contrast, the coefficients
\begin{align}
 \tilde{\Omega}^{\{2\}} \;,\; \tilde{\delta}^{\{2\}} \;,\; \tilde{\gamma}^{\{2\}} \sim 1 
\end{align}
for the second layer are not suppressed due to cavity mode detuning, as they still contain the non-zero coupling coefficient $g^{[3]\{2\}}$ to the resonantly driven mode. From these scalings we can already anticipate the EIT behavior in accordance with the interpretation discussed Ref.~\cite{Roehlsberger2012}:
Only the nuclei in the second layer decay superradiantly. The collective decay of atoms in the first layer $\tilde{\gamma}^{\{1\}}$ and the cross-damping terms $\tilde{\gamma}^{\{1,2\}}$ are quadratically suppressed in the detunings of the additional cavity modes and can be neglected in a first approximation. However, other contributions due to the presence of further cavity modes can have a substantial influence on the system, such as the coherent driving $\tilde{\delta}^{\{1,2\}}$ between the two layers, which can give rise to the coupling field required for EIT.

These scalings with the cavity detuning $\Delta_C$ are visualized in the level scheme shown in Fig.~\ref{fig:level_schemes}(b) as well. Coupling rates denoted by thick, solid or dashed lines and indicate the different powers in the scaling behavior with respect to $\Delta_C$. With the relative magnitude of the rates in mind, a very close similarity with the EIT level scheme from Fig.~\ref{fig:level_schemes}(a) can be observed. Hence, our analysis so far also suggests EIT-like features in the system. However, it is yet unclear how the additional driving terms and inter-layer coupling terms affect the spectrum in detail. In order to answer this question, we will now turn to the analytic solution of the model.

Starting from Eqs.~(\ref{eq:multi_eit_H}) and~(\ref{eq:multi_eit_L}), we find that the equations of motion for the density matrix elements
\begin{align}
 \rho_{1G} &= \bra{E^{\{1\}}}\rho\ket{G}   \;, \\
 \rho_{2G} &= \bra{E^{\{2\}}}\rho\ket{G}  \;,
\end{align}
form a closed set of equations in the limit of linear response, i.e., ~where the populations $\bra{G}\rho\ket{G}\approx 1$ and $\bra{E^{\{1\}}}\rho\ket{E^{\{1\}}} = \bra{E^{\{2\}}}\rho\ket{E^{\{2\}}} \approx 0$ and the coherence between the excited states $\bra{E^{\{1\}}}\rho\ket{E^{\{2\}}}$ vanishes. The equations of motion read
\begin{align}
 \frac{d}{dt}\rho_{1G} &= \left[ i(\Delta-\tilde{\delta}^{\{1\}}) - \tilde{\gamma}^{\{1\}} - \tfrac{\gamma}{2}  \right] \rho_{1G} \nonumber \\ & - i\tilde{\Omega}^{\{1\}} - (i\tilde{\delta}^{\{1,2\}}+\tilde{\gamma}^{\{1,2\}}) \rho_{2G} \;,\\
 \frac{d}{dt}\rho_{2G} &= \left[ i(\Delta-\tilde{\delta}^{\{2\}}) - \tilde{\gamma}^{\{2\}} - \tfrac{\gamma}{2}  \right] \rho_{2G} \nonumber \\ &- i\tilde{\Omega}^{\{2\}}  - (i
{{}\tilde{\delta}^{\{1,2\}}}^*
+{{}\tilde{\gamma}^{\{1,2\}}}^*
) \rho_{1G} \;.
\end{align}
From this we obtain the steady state solutions of the coherences
\begin{align}
 \rho_{1G} &= \frac{\tilde{\Delta}^{\{2\}}\tilde{\Omega}^{\{1\}} - \left( -\tilde{\delta}^{\{1,2\}} + i \tilde{\gamma}^{\{1,2\}} \right) \tilde{\Omega}^{\{2\}} }
{
\tilde{\Delta}^{\{1\}}
\tilde{\Delta}^{\{2\}}
-
\Omega_C^2
} \label{eq:multi_eit_rho1G} \;,\\
 \rho_{2G} &= \frac{\tilde{\Delta}^{\{1\}}\tilde{\Omega}^{\{2\}} - \left( -{{}\tilde{\delta}^{\{1,2\}}}^* + i {{}\tilde{\gamma}^{\{1,2\}}}^* \right) \tilde{\Omega}^{\{1\}} }
{
\tilde{\Delta}^{\{1\}}
\tilde{\Delta}^{\{2\}}
-
\Omega_C^2
} \label{eq:multi_eit_rho2G} \;,
\end{align}
with the abbreviations
\begin{align}
\tilde{\Delta}^{\{l\}} &= \Delta - \tilde{\delta}^{\{l\}} + i \,(\tfrac{\gamma}{2}+\tilde{\gamma}^{\{l\}})\;, \\
\Omega_C^2 &= \left( \tilde{\delta}^{\{1,2\}}  -i\tilde{\gamma}^{\{1,2\}} \right) 
\left( {{}\tilde{\delta}^{\{1,2\}}}^* -i {{}\tilde{\gamma}^{\{1,2\}}}^*  \right) \;. \label{eq:multi_eit_omega_c}
\end{align}

With the solutions for the coherences at hand, we can now turn to the observable, the complex reflection coefficient $R$. According to Eq.~(\ref{eq:R_lin_B0}), it is given by
\begin{align}
 R = -1 + \sum_j \frac{{2 \kappa_R^{[j]}}}{\kappa^{[j]}+i \Delta_C^{[j]}} + R^{\{1\}} \rho_{1G} + R^{\{2\}} \rho_{2G}\;, \label{eq:eit_R}
\end{align}
with
\begin{align}
 R^{\{l\}} = -\frac{i}{\ain} \sum_j  \frac{ \sqrt{2 \kappa_R^{[j]} } }{ \kappa^{[j]}+i \Delta_C^{[j]} }  \sqrt{\tfrac{2}{3}} {g^{[j]\{l\}}}^* \sqrt{N^{\{l\}}} \;. \label{eq:multi_eit_Rl_def}
\end{align}
At this point it is instructive to discuss the scaling related to the cavity detuning $\Delta_C$ once again. As before, we assume that $g^{[3]\{1\}} = 0$, i.e.~the first layer does not couple to the driven cavity mode $j=3$ since it is located at a field node. In this case we find that $R^{\{1\}} \sim 1/\Delta_C$, while $R^{\{2\}}$ is not suppressed due to a cavity detuning, since the second layer can couple to the resonantly driven mode as $g^{[3]\{2\}} \neq 0$. Furthermore, from Eqs.~(\ref{eq:multi_eit_rho1G}) and~(\ref{eq:multi_eit_rho2G}) we find that $\rho_{1G} \sim 1/\Delta_C$, whereas the $\rho_{2G}$ is not suppressed in this fashion. Therefore, for a qualitative understanding of the reflectance, it is well justified to drop the quadratically suppressed contribution $R^{\{1\}} \rho_{1G}$ and only consider the reflection signal which stems from the second layer, i.e.~$R^{\{2\}} \rho_{2G}$.

In a further step, we restrict the numerator of the fraction in $R^{\{2\}} \rho_{2G}$ to terms up to linear order in $1/\Delta_C$. Moreover, we neglect the tiny collective Lamb shift and superradiance of the nuclei in the first layer. This yields the reflection coefficient
\begin{align}
 R = &-1 + \sum_j \frac{{2 \kappa_R^{[j]}}}{\kappa^{[j]}+i \Delta_C^{[j]}} \nonumber \\
  &+ R^{\{2\}} \, \tilde{\Omega}_2 \; \frac{ \Delta + i \,\tfrac{\gamma}{2}}
{
\left( \Delta  + i \, \tfrac{\gamma}{2} \right)
\left( \Delta - \tilde{\delta}_2 + i \,(\tfrac{\gamma}{2}+\tilde{\gamma}_2) \right)
- \Omega_C^2 } \;. \label{eq:multi_eit_R}
\end{align}
The nuclear contribution to the reflectance is revealed in the second line. Its spectral shape is essentially that of a system featuring EIT. Hence, we recover the same result as in Ref.~\cite{Roehlsberger2012}: In a cavity with two resonant layers it is possible to realize the phenomenon of electromagnetically induced transparency.

\subsection{Comparison to the semiclassical analysis\label{sec:comparison}}

In Ref.~\cite{Roehlsberger2012}, the case was studied, where the empty-cavity contribution to the reflectance vanishes, and
a semiclassical theory based on transfer matrix techniques was used to derive an expression for the nuclear reflectance. In the sign convention of the nuclear resonances of the present work, the result was found as
\begin{align}
 R &= - i d_2 f_0 \tfrac{\gamma}{2} E_{2-+} \frac{ \Delta + i\tfrac{\gamma}{2}} {(\Delta + i\tfrac{\gamma}{2}) (\Delta + i\tfrac{\Gamma}{2} ) - \Omega_C^2} \;,\\
 \Gamma &= \gamma (1+d_2 f_0 E_{2--}) \;, \\
 \Omega_C^2 &= d_1 d_2 f_0^2 \tfrac{\gamma^2}{4} E_{2-+}E_{1+-} \;, \label{eq:Omega_C_semi}
\end{align}
where $d_1$ and $d_2$ are the thicknesses of the respective two layers, $f_0$ is the nuclear scattering amplitude at resonance, and $E_{2-+}$, $E_{2--}$, and $E_{1+-}$ are transfer matrix elements. Comparing it with the part of the nuclear reflection in the quantum optical expression given in Eq.~(\ref{eq:multi_eit_R}), we notice a perfect agreement of the structures of the two formulas.
However, as an important consistency check, it remains to be verified if the scaling with the number of nuclei in the two layers agrees as well. In the semiclassical theory it was shown that the amplitude of the reflection coefficient and the superradiance of the nuclei in the second layer scale linearly with the thickness of the second layer $d_2$, and furthermore the control field $\Omega_C$ was shown to be proportional to $\sqrt{d_1 d_2}$. 
The present model does not directly contain the layer thicknesses as parameters. But since $d_1 \propto N^{\{1\}}$ and $d_2 \propto N^{\{2\}}$, is sufficient to show that the scaling relations also hold for the numbers of nuclei. From Eqs.~(\ref{eq:multi_eit_omega_def})--(\ref{eq:multi_eit_g12_def}) and (\ref{eq:multi_eit_Rl_def}) it can indeed be seen that the relations are correctly reproduced by our theory.

This is an important result, since it is a strong hint that the two independently derived results do not coincide by chance, but also agree on a more fundamental level. Hence, the model developed here can now be employed to shine light on the EIT scenario from a different perspective.

In the nuclear reflectance calculated in Eq.~(\ref{eq:multi_eit_R}), the coupling Rabi frequency occurs as $\Omega_C^2$ in the denominator, whereas in standard EIT settings it appears as a positive real-valued variable $|\Omega_C|^2$. Taking a closer look at our definition of the coupling Rabi frequency in Eq.~(\ref{eq:multi_eit_omega_c}), we note that in our case $\Omega_C^2$ can generally be complex. Also in the semiclassical theory the complex field amplitudes and transfer matrix elements $E_{2-+}E_{1+-}$ allow for complex values, c.f.~Eq.~(\ref{eq:Omega_C_semi}). The results of Ref.~\cite{Roehlsberger2012}, though, seem to imply that the imaginary component is very small and an EIT situation is well realized. However, from the theoretical analysis of the semiclassical models, this fact could not be understood and the influence of the imaginary component was unclear~\cite{Roehlsberger2013}. With the present theory, though, it is now possible to examine the complex nature of the coupling in more detail. From Eq.~(\ref{eq:multi_eit_omega_c}) we know that it is not only given by the coherent coupling $\tilde{\delta}^{\{1,2\}}$ between the two layers as written in the Hamiltonian in Eq.~(\ref{eq:multi_eit_H}), but is also affected by the incoherent cross-damping term $\tilde{\gamma}^{\{1,2\}}$ between the two layers. In the discussion on the scalings we have already seen that, in contrast to the coherent contribution, the incoherent term is suppressed quadratically with the detuning of the off-resonant cavity modes. Thus, we find that the  incoherent part
\begin{align}
 \operatorname{Im}(\Omega_C^2) = -2 \operatorname{Re}\left(  \tilde{\delta}^{\{1,2\}} {{}\tilde{\gamma}^{\{1,2\}}}^* \right) \sim \frac{1}{\Delta_C^3} \;
\end{align}
can be neglected, and $\Omega_C^2 \approx |\tilde{\delta}^{\{1,2\}}|^2$ such that the real component of the coupling frequency $\Omega_C$ dominates.

Furthermore, the microscopic ansatz of our quantum optical theory enables one to interpret the origin of the coupling between the layers. While in Ref.~\cite{Roehlsberger2012} it was shown that the EIT control field arises from radiative coupling between the two resonant layers, it can now be pinned down from Eqs.~(\ref{eq:multi_eit_d12_def}),~(\ref{eq:multi_eit_g12_def}) and~(\ref{eq:multi_eit_omega_c}) to
\begin{align}
 \Omega_C^2 &= \left(\tfrac{2}{3}\right)^2 N^{\{1\}} N^{\{2\}}
\nonumber \\ &\quad\times
\left( \sum_j \frac{g^{[j]\{1\}} {g^{[j]\{2\}}}^*}{\Delta_C^{[j]}-i \kappa^{[j]}} \right) 
\left( \sum_j \frac{{g^{[j]\{1\}}}^* {g^{[j]\{2\}}}}{\Delta_C^{[j]}-i \kappa^{[j]}} \right) \;.
\end{align}
Since we assumed that the first layer does not couple to the third guided mode in the idealized case, i.e.~$g^{[3]\{1\}} = 0$, we observe that the coupling field is only mediated via the remaining guided modes $j\neq 3$ in the cavity. This way, it becomes now also clear why the EIT phenomenon was not obtained in Sec.~\ref{sec:effect_layers}, where multiple layers, but only one guided mode was included in the theoretical analysis.

\subsection{Numerical analysis\label{sec:numerical_analysis}}

\begin{figure*}[!t]
 \centering
 \includegraphics[scale=1]{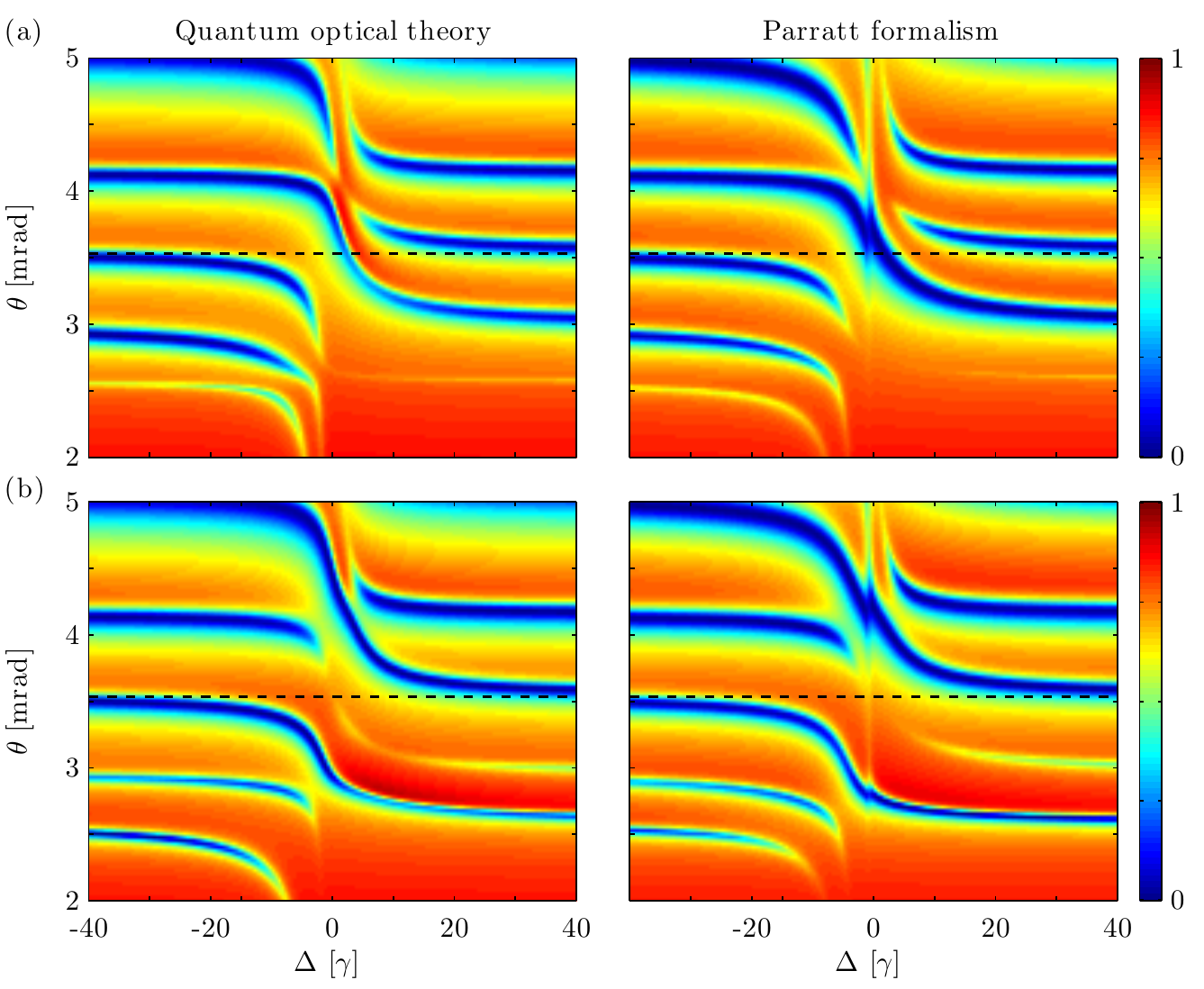}
 \caption{(Color online) The reflectance of (a) the EIT and (b) the non-EIT scenario is shown as a function of the detuning $\Delta$ and the incidence angle $\theta$. The results derived with the extended quantum optical model agree very well with the predictions from Parratt's formalism. The dashed line at $\theta \approx 3.5$ mrad marks the angle at which the 3rd cavity minimum is expected. A cut along this line corresponds to the spectrum measured in Ref.~\cite{Roehlsberger2012} and is shown in Fig.~\ref{fig:eit_qo_parratt_spectrum}. Parameters are given in Appendix~\ref{sec:appendix_parameters}.
\label{fig:eit_noneit_qo_parratt_grid}
}
\end{figure*}

Let us now see how well our analytical expression for the reflectance derived above performs in practice. In particular, we aim to describe the spectrum of the EIT cavity defined in the first column of Tab.~\ref{tab:layers_eit} with our quantum optical model. Moreover, we include a second cavity into the analysis: While the EIT cavity has its resonant layers in a node and antinode of the field of the resonantly driven mode, respectively, we also consider a cavity in which the situation is reversed. Namely, the first resonant layer is located at a field antinode and the second ensemble of nuclei at the field node. The corresponding geometry is defined in the last column of Tab.~\ref{tab:layers_eit}. The two cavity layouts reflect the cases discussed in Ref.~\cite{Roehlsberger2012}, where it was shown that the first cavity exhibits the EIT phenomenon, while for the second system the control coupling $\Omega_C$ vanishes and only a Lorentz-like spectrum is measured.

In order to determine the free parameters related to the cavities defined in Tab.~\ref{tab:layers_eit} for the quantum optical model in a consistent way, we employed the following method. First, we restricted ourselves to the first five guided modes in the theory and did not take into account the resonant nuclei yet. For each of these modes the angles $\theta_0^{[j]}$, at which the modes are driven resonantly, and the decay and coupling rates $\kappa^{[j]}$ and $\kappa_R^{[j]}$ have to be determined. The parameters can be found by fitting Eq.~(\ref{eq:R_C_multimode_heuristic}) to the reflection curve as function of the x-ray incidence angle as it was already done in Sec.~\ref{sec:effect_modes}.

With the cavity parameters at hand, the next step now is to include the nuclear resonances to the model. In particular, the complex collective coupling coefficients $g^{[j]\{l\}}\sqrt{N^{\{l\}}}$ between the $j$th guided mode and the layer $l$ of resonant nuclei have to be determined. For each cavity the number of coupling coefficients is $10$, since we specialize to five guided modes in the analysis and each mode can couple to the two respective layers. In order to avoid arbitrariness in a fit to numerical data, it is advisable to decrease this large number of free parameters. In fact, is it possible to determine all coupling coefficients in a consistent way, while keeping only one global scaling as free parameter. To illustrate this, we note that the coupling coefficients can be decomposed as
\begin{align}
 g^{[j]\{l\}}\sqrt{N^{\{l\}}} = \tilde{\mathcal{E}}^{[j]\{l\}} \; \cdot \; \big( \tilde{g}^{\{l\}}\sqrt{N^{\{l\}}} \big) \;, \label{eq:g_decomposed}
\end{align}
where the first factor denotes the cavity field amplitude of mode $j$ at layer $l$, and the second factor includes the collective nuclear dipole moment. Next, we exploit that the complex field amplitudes in the cavity can be easily derived by means of Parratt's formalism~\cite{deBoer1991}.
In a simple picture, we can interpret the resonant nuclei in the cavity as a perturbation, which modifies the cavity field and, accordingly, the reflectance. The cavity field in the presence of nuclear resonances can be understood as a superposition of the bare cavity field and the contribution due to scattering at the nuclei. This presupposition clearly holds for x-ray frequencies apart from the nuclear resonance, but does also give consistent results directly at the resonance, where the perturbation due to the nuclei is not generally small. Hence, to determine the field coefficients $\tilde{\mathcal{E}}^{[j]\{l\}}$, it is not necessary to include any nuclear resonances in Parratt's formalism, but only the bare cavity field in the absence of $^{57}$Fe resonances are required. With the input field normalized to intensity one, like in Fig.~\ref{fig:field}, we can directly compute all complex valued field coefficients $\tilde{\mathcal{E}}^{[j]\{l\}}$ at the center of the respective layers $l$ by tuning the incidence angle $\theta$ to the angles $\theta^{[j]}$, where the $j$th cavity mode is driven resonantly. The remaining task is to determine the second coefficient in Eq.~(\ref{eq:g_decomposed}), which takes into account the nuclear properties and other constant contributions. Since both iron layers in the cavities have the same thickness and hence the number of nuclei is the same, we can expect that $\tilde{g}^{\{l\}}\sqrt{N^{\{l\}}}$ is a constant and acts only as a scaling parameter for the previously determined field amplitudes. This way, all coupling coefficients $g^{[j]\{l\}}\sqrt{N^{\{l\}}}$ can be deduced by fitting the model to numerical data, calculated with Parratt's formalism, with only one free scaling parameter. The complex field amplitudes for both cavities and the scaling parameter found in our analysis are summarized in Appendix \ref{sec:appendix_parameters}. We note that the couplings to the layers, which are located in the field nodes of the third cavity mode, do not completely vanish due to the finite thickness of the layers and a potential misplacement in the cavity. However, they are found to be much smaller than the coupling coefficients of the respective layers in the cavity field antinode. This can already be deduced from the field intensity distributions shown in Fig.~\ref{fig:field}.

An alternative approach to determine the coupling constants $g^{[j]\{l\}}\sqrt{N^{\{l\}}}$ is by fitting the model with all coefficients directly to numerical data. While this procedure is not as persuasive as the consistent method described above, it might also offer some advantages in quantitative studies. Errors in other parameters, such as the coefficient which characterizes the cavity modes can partly be compensated. Moreover, for iron layers with a larger thickness the field amplitude might not be constant and, in contrast to the method from above, an effective coupling strength would be naturally obtained. Finally, the fitted parameters could provide a handle to cover the fact in more detail, that on-resonance the nuclei have an effect on the cavity field which goes beyond a perturbation. In this work, however, we will not optimize the parameters in this way, but utilize the coefficients derived previously to illustrate the general consistency of our model.

Now we are able to benchmark our analytical result for the case of two resonant layers, which was calculated in Eqs.~(\ref{eq:multi_eit_rho1G})--(\ref{eq:multi_eit_Rl_def}). A comparison with the frequency- and angular-dependent reflectance for the EIT and the non-EIT cavity is shown in Fig.~\ref{fig:eit_noneit_qo_parratt_grid}. Clearly, the agreement between the two different models is very good. We stress that this is not an obvious result, since the parameters for the quantum optical model were determined independently and not obtained from a fit to the numerical data.

A range, in which strong deviations can be observed, is the domain around $\Delta\approx 0$. Here, the exact numerical solution obtained from Parratt's formalism shows an additional structure. This can be understood from the following considerations. If the x rays are not resonant to the transition in the $^{57}$Fe nuclei, they will primarily be damped due to the electronic absorption in the cavity, before they can reach the lower resonant layer. If, however, their frequency is too close to resonance, the x rays will additionally be absorbed by the nuclei in the upper layer. Consequently, the field seen by the nuclei in the second layer is strongly modified compared to the off-resonant case. However, in the derivation above we assumed that the presence of the nuclei can be treated as a small perturbation to the cavity field, which is not the case in the extreme situation encountered here. An approach for future studies could thus be to comprise this effect self-consistently into the quantum optical theory for an even better agreement with the numerical data.

\begin{figure}[t]
 \centering
 \includegraphics[scale=1]{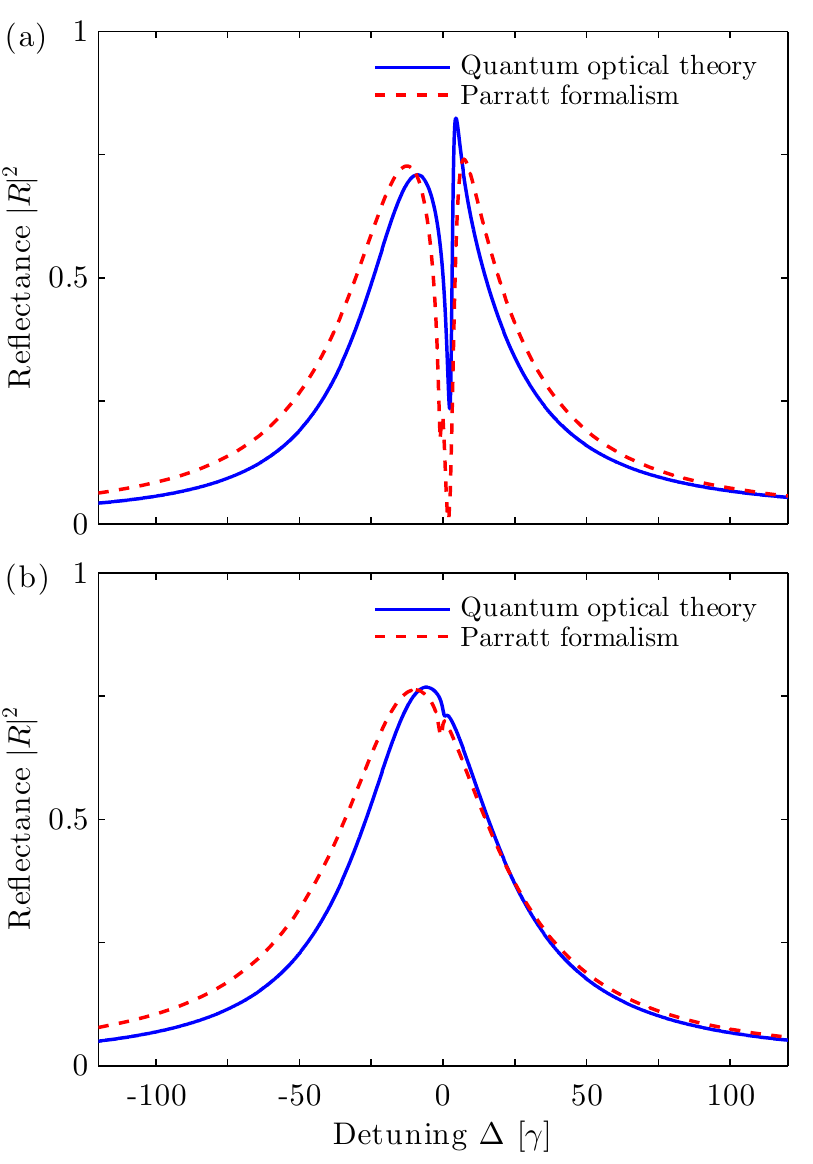}
 \caption{\label{fig:eit_qo_parratt_spectrum}(Color online) Spectra of (a) the EIT and (b) the non-EIT cavity at incidence angle $\theta \approx 3.5$~mrad, at which the third guided mode is excited. The quantum optical description (blue solid lines) is in qualitative agreement with the exact result derived with the Parratt formalism (red dashed lines). Parameters are given in Appendix~\ref{sec:appendix_parameters}.}
\end{figure}

We now turn to the spectrum measured at the incidence angle corresponding to the third guided mode, i.e., the situation from Ref.~\cite{Roehlsberger2012}. The spectra for both the EIT and the non-EIT cavity defined in Tab.~\ref{tab:layers_eit} are shown in Fig.~\ref{fig:eit_qo_parratt_spectrum}.
Again, we observe a good qualitative agreement of our theory with the numerical data obtained with Parratt's formalism, which could already be anticipated from the accordance in Fig.~\ref{fig:eit_noneit_qo_parratt_grid}. But in any case, the fact that the EIT as well as the non-EIT spectrum is reproduced without post-optimization of the consistently derived parameters, supports the validity of our theoretical description.

Finally, we want to review to role of the coupling field $\Omega_C$. In the theoretical analysis in Sec.~\ref{sec:comparison} it was found that the presence of this control field gives rise to EIT. Moreover, we found that the control field is established by an interaction between the nuclear ensembles via different cavity modes, as the layer in the field node does not directly couple to the driven guided mode. From our numerical analysis we observe that this idealized case is not strictly realized. Since the coupling coefficient is small yet finite,  also the resonantly driven mode gives rise a coupling between the two layers in the cavity. Furthermore, the control field $\Omega_C$ does not vanish in the non-EIT case and hence the Lorentz-like spectrum cannot be explained by its absence in the frame of our model. Rather, in the non-idealized case it is the interplay with other contributions to the reflection coefficient and their interference which results in the Lorentzian spectrum.

\section*{Summary and discussion}

In summary, we investigated the effect of multiple modes and multiple ensembles of resonant M\"ossbauer nuclei in an x-ray cavity QED setup, which has recently served as a platform for multiple experiments related to x-ray quantum optics.
Most of the time, the scenario with a singe ensemble, realized by a layer of collectively acting nuclei, has been studied theoretically~\cite{Heeg2013b} as well as in several experiments~\cite{Roehlsberger2010,Heeg2013,Heeg2014,Heeg2014b}.
The theoretical framework of this work is applicable to also model experimental settings with more than one resonant ensemble~\cite{Roehlsberger2012} and interpret them from a quantum optical point of view.

Our theory from Sec.~\ref{sec:generalization} is based on the approach taken in Ref.~\cite{Heeg2013b} and constitutes a generalization with multiple cavity modes and several layers of resonant M\"ossbauer nuclei. Similar to the original theory, we were able to simplify the basic equations using two well justified approximations. By adiabatically eliminating the cavity modes and restricting the analysis to the linear regime, effective equations of motion for the nuclear ensembles could be derived. The resulting set of equations characterizing the dynamics of a few-level system can easily be solved analytically.

In Sections \ref{sec:effect_modes} and \ref{sec:effect_layers} we discussed the consequences of the two extensions to the theory in detail. By introducing multiple cavity modes to the model we found that the spectral properties around the resonance of $^{57}$Fe are unaffected. In absence of magnetic hyperfine splitting, the nuclear response is given by Lorentz profiles, which are shifted and broadened due to collective effects. The differences to the predictions from Ref.~\cite{Heeg2013b} manifest only in the coefficients entering the final expressions. However, a clear difference could be observed when the reflectance was studied as a function of the x-ray incidence angle. While a single-mode theory can only indicate one guided mode of the system at a time, our extension allows to accurately model the reflectance over a range of several mrad, reproducing all guided modes. Moreover, we found that our model, which takes into account the effect of the cavity and its modes, can be heuristically extended to incorporate bulk material properties such as the total reflection envelope. This way, a close agreement with established semiclassical models could be achieved.
Further, the effect of multiple ensembles of $^{57}$Fe nuclei in the cavity, located in different layers, was studied. This extension alone did not give rise to qualitatively new effects.

Next, we analyzed the case in which both extensions enter the theory at the same time, i.e.~multiple cavity modes and multiple resonant layers. We could show that in this case the equations cannot be mapped to an effective two-level system, as the coupling coefficients between the different nuclear ensembles and cavity modes are mutually different and do not allow for a diagonalization in which only one excited state is probed. Rather, more advanced level schemes generally occur in this setting.

In the final part of this work we applied the general theory to the setting, which was experimentally explored in Ref.~\cite{Roehlsberger2012}. In this reference, EIT-like spectra could be observed for a cavity with two layers of resonant iron nuclei. We applied our quantum theoretical approach to the setting and could successfully reproduce the findings. An effective level scheme with one collective ground and two collective excited states which captures the complete system could be found and an analytic solution for the reflection coefficient was given. For the idealized case of perfectly placed layers in the cavity we found that the nuclear response has indeed the spectral shape of a system featuring EIT. In this process, we compared our result to the previously used semiclassical models and observed agreement on the analytic level. In particular, the scalings with respect to the number of atoms in the respective ensembles are reproduced by our quantum optical description. Most importantly, the question on the nature of the control field, which forms a pivotal requirement of EIT, could be elucidated. From our analysis of the idealized scenario we found that the radiative coupling is mediated by the off-resonant cavity modes.

We further developed a way to consistently derive the different coupling rates required for the model. This approach is based on an analysis of the cavity in the absence of nuclear resonances. Hence, the arising spectral features can be traced back to the capability of our model and are not due to a potentially biased parameter fit. In our numerical data we observe a good agreement to the results of semiclassical models and the essential features, such as the signatures of EIT, are reproduced.

While we mainly analyzed the cavity properties in the absence of magnetization, we emphasize that the extended theory description developed in this work is not restricted to a vanishing magnetic hyperfine splitting in the resonant layers. Rather, in our model it is possible to include all Zeeman sublevels properly. In future works, this could be exploited to combine the effect of multiple layers and modes, giving rise to the EIT-like effects, and magnetization, leading to the phenomenon of spontaneously generated coherences~\cite{Heeg2013}. This way, a broad class of quantum optical level schemes could be engineered, indicating promising perspectives of x-ray cavity QED with M\"ossbauer nuclei.

\section*{Acknowledgements}
Fruitful discussions with R. R\"ohlsberger are gratefully acknowledged. K.P.H. acknowledges funding by the German National Academic Foundation.

\appendix
\section{Reflection curve envelope due to bulk material properties\label{sec:appendix_envelope}}
As explained in the main text, the quantum optical theory can model the minima in the reflection curve $R(\theta)$ indicating the guided modes of the cavity, whereas the envelope formed by the total reflection behavior is not part of the description (gray dotted line in Fig.~\ref{fig:eit_rocking_qo_parratt}). However, it can be included by combining the expressions of the quantum optical model, describing the cavity structure, with an envelope function $R_\textrm{Envelope}(\theta)$, which takes into account the reflection of the bulk material. This way, a good agreement to the semiclassical approaches is achieved. For a single (infinitely thick) layer, the Fresnel reflection coefficient reads
\begin{align}
 R_\textrm{Envelope}(\theta) = \frac{ \sin{(\theta)} - \sqrt{\sin{(\theta)}^2 + n^2 - 1} }{ \sin{(\theta)} + \sqrt{\sin{(\theta)}^2 + n^2 - 1} } \;,
\end{align}
where $\theta$ is the angle of incidence and $n$ is the refractive index of the material. For a platinum layer and at x-ray energy $14.4$~keV, the latter is given by~\cite{Schoonjans2011}
\begin{align}
 n &= 1 - \delta + i \beta \;,\\
 \delta &= 1.603365 \times 10^{-5} \;,\\
 \beta &=  2.56353  \times 10^{-6} \;.
\end{align}

\section{Numerical parameters\label{sec:appendix_parameters}}
In Tabs.~\ref{tab:parameters1} and~\ref{tab:parameters2} we summarize the parameters for our quantum optical theory, which we used in this work.
\begin{widetext}

\begin{table*}[!h]
 \centering
 \caption{\label{tab:parameters1}{Parameters for the EIT cavity.}}
\begin{tabular}{c @{\hspace{0.3cm}} r @{\hspace{0.3cm}} r @{\hspace{0.3cm}} r @{\hspace{0.5cm}} r @{$\:$}r@{$\:i$} p{0.3cm} @{\hspace{0.3cm}} r @{$\:$}r@{$\:i$}  p{2.2cm}} 
 \hline \hline \\[-1.5ex]
 Mode $j$ & $~~\theta_0^{[j]}$ [mrad] & $\kappa^{[j]}~[\gamma]~~$ & $\kappa_R^{[j]}~[\gamma]~$ &
\multicolumn{3}{c}{$\tilde{\mathcal{E}}^{[j]\{1\}}$\hspace*{4mm}} &
\multicolumn{3}{c}{\hspace*{7mm}$\tilde{\mathcal{E}}^{[j]\{2\}}$\hspace*{10mm}} \\[0.5ex]
 \hline \\[-1.5ex]
1 & 2.55943 & 145807  & 6110   & $ 0.609 \: +$ & $ 0.036$ & & $ 0.903 \:+$ & $ 0.346 $ \\
2 & 2.99211 & 533322  & 311376 & $ 2.105 \: +$ & $ 1.795$ & & $ 0.818 \:+$ & $ 0.810 $ \\
3 & 3.54936 & 615909  & 275736 & $-0.031 \: +$ & $ 0.440$ & & $-1.683 \:-$ & $ 1.815 $ \\
4 & 4.14850 & 783648  & 373532 & $-0.947 \: -$ & $ 1.130$ & & $-0.244 \:-$ & $ 0.781 $ \\
5 & 5.07939 & 1718031 & 767833 & $-0.370 \: -$ & $ 1.567$ & & $ 0.361 \:+$ & $ 1.582 $ \\[0.5ex]
  \hline \hline
\end{tabular}
\end{table*}

\begin{table*}[!ht]
 \centering
 \caption{\label{tab:parameters2}{Parameters for the non-EIT cavity.}}
\begin{tabular}{c @{\hspace{0.3cm}} r @{\hspace{0.3cm}} r @{\hspace{0.3cm}} r @{\hspace{0.5cm}} r @{$\:$}r@{$\:i$} p{0.3cm} @{\hspace{0.3cm}} r @{$\:$}r@{$\:i$}  p{2.2cm}} 
 \hline \hline \\[-1.5ex]
 Mode $j$ & $~~\theta_0^{[j]}$ [mrad] & $\kappa^{[j]}~[\gamma]~~$ & $\kappa_R^{[j]}~[\gamma]~$ &
\multicolumn{3}{c}{$\tilde{\mathcal{E}}^{[j]\{1\}}$\hspace*{4mm}} &
\multicolumn{3}{c}{\hspace*{7mm}$\tilde{\mathcal{E}}^{[j]\{2\}}$\hspace*{10mm}} \\[0.5ex]
 \hline \\[-1.5ex]
1 & 2.58446 & 242554  & 150696   & $ 1.889 \: +$ & $ 1.613 $ & & $ 0.957 \:+$ & $ 0.935 $ \\
2 & 2.96021 & 311202  & 48680 & $-0.760 \: -$ & $ 0.140 $ & & $-1.648 \:-$ & $ 1.160 $ \\
3 & 3.55108 & 607732  & 342667 & $-1.631 \: -$ & $ 2.086 $ & & $-0.058 \:-$ & $ 0.332 $ \\
4 & 4.17107 & 922367  & 453227 & $ 0.203 \: -$ & $ 0.067 $ & & $ 1.031 \:+$ & $ 1.606 $ \\
5 & 5.09251 & 1844998 & 833796 & $ 0.217 \: +$ & $ 1.585 $ & & $-0.203 \:-$ & $ 1.604 $ \\[0.5ex]
  \hline \hline
\end{tabular}
\end{table*}

\end{widetext}
Note that the coefficients
$\tilde{\mathcal{E}}^{[3]\{1\}}$ for the EIT cavity and
$\tilde{\mathcal{E}}^{[3]\{2\}}$ for the non-EIT cavity
are near zero, indicative of a field node at the location of the respective layers.
The asymmetry parameters $r$ for the two respective cavities, expected to be close to the value $-1$, were determined as
$r_\textrm{EIT} \approx  -0.981 + 0.363 \: i \approx -1.047 \: e^{- 0.354 i}$
and
$r_\textrm{non-EIT} \approx -0.979 + 0.383\: i \approx -1.051\:e^{-0.373 i}$. For the scaling parameter of the coupling coefficients the values
$\tilde{g}^{\{1\}}\sqrt{N^{\{1\}}} = \tilde{g}^{\{2\}}\sqrt{N^{\{2\}}} = 1983.89 \,\gamma$
were used.

\bibliography{multi}
\end{document}